\documentclass[conference]{IEEEtran}
\IEEEoverridecommandlockouts
\usepackage[T1]{fontenc}
\usepackage[utf8]{inputenc}
\usepackage{textcomp}
\usepackage{graphicx}
\usepackage{booktabs}
\usepackage{array}
\usepackage{caption}

\usepackage{url}
\usepackage[hidelinks]{hyperref}
\usepackage{balance}
\usepackage{xcolor}
\usepackage{needspace}
\usepackage{listings}
\begin{document}

\title{A Security Analysis of Verifiable Credential Verification Pipelines:\\ Attack Surface Arising from JSON-LD/RDFC Canonicalization\\ Compared with SD-JWT VC and mdoc}

\author{\IEEEauthorblockN{Naohiro Fujie}
\IEEEauthorblockA{\textit{MIRAI Design Laboratory}\\
\textit{ITOCHU Techno-Solutions Corporation}\\
\textit{OpenID Foundation Japan}\\
Tokyo, Japan\\
naohiro.fujie@ctc-g.co.jp}
\and
\IEEEauthorblockN{Shigeya Suzuki}
\IEEEauthorblockA{\textit{Global Research Institute}\\
\textit{Keio University}\\
Tokyo, Japan\\
shigeya@wide.ad.jp}}

\maketitle

\thispagestyle{plain}
\pagestyle{plain}

\begin{abstract}
We evaluate, at implementation level and under an explicit threat model, the verification pipelines of three Verifiable Credential (VC) formats: SD-JWT VC; the W3C Verifiable Credentials Data Model 2.0 with the Data Integrity cryptosuite \texttt{eddsa-rdfc-2022} (W3C VCDM); and mdoc (ISO/IEC 18013-5). For W3C VCDM we identify and exercise three structural attack surfaces from remote context resolution and RDFC-1.0 canonicalization: denial of service (DoS) via poison graphs, context injection through \texttt{@context}, and server-side request forgery (SSRF) reachable via malicious URLs, all at default settings. A poison graph, whose blank-node references are symmetric enough to force RDFC-1.0 into deep recursion, raises it to 278$\times$ the baseline from five blank nodes in 527 bytes, and to 13.5 s from eight. Every experiment ran in four environments sharing runtime and library versions; the multiple moves with the machine (251--424$\times$ for one input), not with its speed. Cost tracks that symmetry rather than the node count: an acyclic input with 1,500 blank nodes stays at 462--714$\times$. A call limit separates the two: every attack family aborts at 16 or below, every legitimate credential completes at 1, but a shallow, wide recursion, such as the acyclic control, passes whatever the bound. DoS does not presuppose adversarial input: with production education schemas, canonicalization costs 23--34$\times$ a simple credential in one environment, 16--49$\times$ across four. Effective mitigations are a static document loader, which removes the retrieval latency and both remote-context surfaces; a context allowlist; and a call limit with a size bound. SD-JWT VC and mdoc need neither canonicalization nor external references, and our verifiers resisted every vector evaluated. These are not format-specific defects but properties of the pipeline and its configuration.
\end{abstract}


\section{Introduction}

In the field of digital identity, Verifiable Credentials (VCs) have attracted attention as a mechanism that lets a verifier mechanically verify issuer-signed claims without querying a third party. The principal options available to implementers today are SD-JWT VC \cite{ref1}, configurations based on the W3C Verifiable Credentials Data Model 2.0 (VCDM 2.0) \cite{ref2}, and mdoc \cite{ref3}. These are not specifications at the same level of abstraction. SD-JWT VC and mdoc are credential formats that jointly specify claim structure, encoding and securing. VCDM 2.0, by contrast, specifies only an abstract data model for credentials; securing is supplied by separately combining it with a securing mechanism (Data Integrity \cite{ref4} or VC-JOSE-COSE \cite{ref5}). This paper evaluates the configuration that combines VCDM 2.0 with the Data Integrity cryptosuite eddsa-rdfc-2022; we refer to it as ``VCDM 2.0 + Data Integrity'' and abbreviate it as W3C VCDM. Where we mean the data model specification alone, without a securing mechanism, we write ``VCDM 2.0'', to distinguish it from the configuration evaluated here. The three rest on different design philosophies, so selecting among them requires multifaceted evaluation.

SD-JWT VC is specified as an IETF Internet-Draft under active development and adopts JWT compact serialization. It has strong affinity with the existing OAuth 2.0/OpenID Connect ecosystem, and selective disclosure is realized through the SD-JWT specification \cite{ref6}. mdoc is the format for mobile identification documents specified in ISO/IEC 18013-5 \cite{ref3} and adopts binary encoding based on CBOR (Concise Binary Object Representation) and COSE (CBOR Object Signing and Encryption). VCDM 2.0 + Data Integrity, by contrast, expresses the credential data model in JSON-LD, and the eddsa-rdfc-2022 cryptosuite requires RDF dataset canonicalization (RDFC-1.0, the W3C Recommendation of the algorithm formerly developed as URDNA2015) \cite{ref7} and the resolution of external contexts for signing and verification.

Studies that compare the security characteristics of these three verification pipelines at the implementation level, under an explicit threat model, remain limited. W3C VCDM in particular requires RDF canonicalization and the resolution of external contexts for signing and verification, so it contains processing stages---absent from the other two formats---that can become an attack surface. The W3C RDFC-1.0 specification \cite{ref8} documents its worst-case complexity and the associated denial-of-service concern, but the extent to which this materializes in implementations, and how effective each mitigation is, have not been shown quantitatively. Moreover, EU Regulation 2024/1183 (eIDAS 2.0) \cite{ref9} establishes the European Digital Identity Wallet (EUDIW), and the implementing acts adopted under it designate SD-JWT VC and mdoc as the credential formats wallets must support, making the comparison of security characteristics across formats increasingly relevant in practice. This paper evaluates, at the implementation level and under an explicit threat model, the resilience of each format against attack vectors including denial of service, context injection, SSRF, algorithm confusion, and data tampering, and quantifies the effectiveness of the available mitigations.

\section{Related Work}

Several prior studies address VC formats, but work that evaluates the attack surface arising from the verification pipeline at the implementation level, under an explicit threat model, remains limited. This section surveys the major related work and clarifies the positioning of this paper.

Brunner et al. \cite{ref10} comprehensively surveyed the technical landscape of Decentralized Identifiers (DIDs) and VCs. After organizing the workflows of issuer, holder, and verifier in accordance with the standards, they discuss the open problems of DIDs and VCs from four perspectives: trust, privacy, revocation, and usability. On trust, they observe that although possession of a DID can be proven cryptographically, binding it to a real-world identity is outside the scope of the standard, so either a web of trust among public issuers or a hybrid arrangement with certificate authorities is required. On revocation and key management, they note the absence of a standard mechanism for proving the time of revocation and the lack of any recovery path when a private key is lost. Their security discussion is thus a conceptual analysis of the trust model and operation of the DID/VC ecosystem as a whole; it does not address the attack surface specific to credential formats---the processing stages of RDF canonicalization and external context resolution. This paper is complementary in that it opens up the signature verification step of the workflow they describe and evaluates the per-format attack surface at the implementation level under an explicit threat model.

Regarding the comparison of selective disclosure mechanisms, Wu and Tian \cite{ref11} conducted a systematic literature review (SLR) of BBS+ signatures and SD-JWT, screening 226 records published between 2017 and 2025 from IEEE, ACM, SpringerLink, ScienceDirect, IETF, and W3C repositories according to Kitchenham's protocol and analyzing 31 primary studies with empirical data. Their review treats security and privacy guarantees as an explicit evaluation axis: BBS+ offers strong unlinkability, predicate proofs, and zero-knowledge disclosure, whereas SD-JWT integrates readily with existing JOSE/OAuth infrastructure but carries a correlation risk because its salted digests remain stable across presentations. On performance they report that BBS+ derived proofs have a fixed size of roughly 140 bytes and verify in about 12 ms, while SD-JWT presentations grow with the number of disclosed claims yet still verify in under 10 ms when two claims are disclosed; they recommend BBS+ for privacy-critical scenarios and SD-JWT for high-throughput web applications. Their security analysis, however, is confined to the guarantees that selective disclosure primitives achieve as cryptographic schemes (unlinkability, minimal disclosure, predicate proofs), and, by the nature of literature synthesis, it does not expose verification implementations to attack inputs. This paper evaluates experimentally the attack surface arising from the implementation of the verification pipeline (RDF canonicalization and external context resolution), and therefore targets a different layer.

Buldini et al. \cite{ref12} proposed a compact selective disclosure scheme (CSD-JWT) based on cryptographic accumulators, showing up to 46\% lower memory usage and up to 93\% smaller Verifiable Presentations than SD-JWT. Their work states an explicit threat model based on the Dolev--Yao adversary, defines the adversary's goals as disclosing personal information within a VC and impersonating the holder, and discusses resilience against three threats---replay attacks, data overcollection, and a compromised communication channel---together with formal definitions of replay resilience and data minimization; it is therefore prior work that includes a security analysis. In particular, they point out that for SD-JWT the structural metadata of undisclosed claims (their number and position) is exposed at verification time and may enable side-channel inference attacks, and they avoid this by aggregating all claims into a fixed-length accumulator value. Their analysis targets the cryptographic properties of the presentation protocol layer, and their evaluation is limited to a comparison of SD-JWT and CSD-JWT: W3C VCDM and mdoc are out of scope, and the attack surface arising from the implementation of verification processing---canonicalization and external context resolution---is not addressed. This paper complements their work by evaluating that implementation-layer attack surface across all three formats.

For the security of RDF dataset canonicalization, the security considerations of the RDFC-1.0 specification \cite{ref8} officially document its worst-case complexity and denial-of-service exposure. On the theoretical side, Hogan \cite{ref13} shows that labeling blank nodes requires exponential time in the worst case while such hard cases are rare among 9.9 million real-world graphs. Our measurements are the security-relevant corollary of that pair of facts: rarity in observed data says nothing about the cost of constructing a hard case deliberately, and we find that a 527-byte document suffices. This paper evaluates that specification-level concern at the implementation level and provides a quantitative analysis of three attack vectors: the DoS attack surface via poison graphs, context injection, and SSRF.

The W3C Verifiable Credentials Working Group has also published a threat model for the VCDM \cite{ref14} as a Group Note Draft (September 2026). It adopts an analysis framework based on data-flow diagrams and STRIDE (spoofing, tampering, repudiation, information disclosure, denial of service, and elevation of privilege), defines the issuer, holder and verifier stakeholders, and enumerates 25 threats (T1--T25). Most of them concern privacy and operational matters such as correlation, excessive disclosure and key management, and not one threat corresponds to the D (Denial of Service) category of STRIDE: the loss of availability caused by the computational cost of verification, or by delays in retrieving external resources, is not treated as a threat in its own right. The same document nevertheless states the verifier's interest explicitly: verification ``must complete within their operational time budgets even when processing untrusted, and potentially malicious, input.'' This paper complements that document by quantifying, at the implementation level, the concrete attack surface behind that availability concern---the cost of RDFC-1.0 canonicalization and the resolution of external contexts. The context substitution we study corresponds to its T7 (Tampering with Unprotected External Resources), and our mdoc data element and COSE header tampering tests correspond to T1 (Credential Content Tampering).

The attack surfaces examined here belong to a class that recurs whenever a verifier processes an expressive document format, and the paper is best positioned against that lineage rather than against VC-specific work alone. Crosby and Wallach \cite{ref15} introduced algorithmic complexity attacks, in which an adversary supplies input chosen to drive a data structure into its worst case, converting a small message into a large amount of computation; the worst-case behavior of blank node labeling that Hogan \cite{ref13} analyzes is such a case, and the poison graph evaluated in Section~\ref{sec:r-poison} is its instantiation for RDFC-1.0. Sp\"{a}th et al. \cite{ref16} surveyed 30 XML parsers across six languages and found 66\% of the default configurations vulnerable to attacks such as external entity expansion and Billion Laughs; the pairing of a remotely resolved reference with an expensive canonicalization step that we examine in JSON-LD is the same pattern one layer up, and their finding that the exposure lies in default configurations is the premise of our own decision to test library defaults (Section~\ref{sec:threat}). On how worst-case inputs are obtained, Petsios et al. \cite{ref17} showed that they can be discovered by fuzzing guided by resource consumption rather than constructed by hand; we construct ours by hand from graph families chosen for their symmetry, and resource-guided fuzzing remains the natural route to families we did not enumerate. Closest to our tests on SD-JWT VC, Detering et al. \cite{ref18} analyzed JOSE implementations and found severe vulnerabilities in six popular libraries, from signature exclusion to Bleichenbacher attacks. Our contribution relative to that work is not a deeper analysis of JOSE but the extension of the same implementation-level question to three credential formats and, in particular, to the canonicalization layer that JOSE-based formats do not have.

Against this background, the contributions of this paper are threefold. First, we take the security risks of the JSON-LD/RDFC verification pipeline beyond theoretical identification in the specification and reduce them to implementation-level evidence under an explicit threat model. Second, we show that denial of service arises from two independent directions and quantify both: adversarial input, where a 527-byte poison graph costs 278$\times$ the baseline and a 1,196-byte one costs 13.5 seconds, and benign input, where production schemas cost 23--34$\times$ a simple credential in one environment and 16--49$\times$ across four machines with a growth rate that rises with attribute count. Separating the two matters because they call for different controls. Third, we measure the effectiveness of each mitigation---static document loader, context allowlist, and call limit---including where each one fails, and give implementers concrete settings to adopt. This paper does not aim to render a superiority judgment among VC formats; its main purpose is to clarify the attack surface that depends on the verification pipeline and its operational configuration.

\section{Technical Specifications of Each Format}

The three formats are described here only insofar as their structure bears on the attack surfaces evaluated: what the signature covers, whether canonicalization is required, and whether verification resolves anything external. The data models and encodings themselves, with worked structural examples of each format, are given in our performance evaluation of the same three formats \cite{ref19}, and we refer the reader there rather than repeating them.

\subsection{Overview}

Table~\ref{tab:1} shows the basic configuration of the three formats evaluated in this paper. Each format differs in its data representation, securing mechanism, signature algorithm and canonicalization requirement, and these differences directly affect the security properties evaluated here. The following sections detail, for each format, the standards position, the signing input and the verification procedure.

\begin{table*}[!t]
\centering
\caption{Table 1. Basic Configuration of Evaluated Formats}\label{tab:1}
\footnotesize\setlength{\tabcolsep}{4pt}
\begin{tabular}{@{}l p{2.9cm} p{3.4cm} l p{2.3cm} p{3.6cm}@{}}
\toprule

\textbf{Format} & \textbf{Data Representation} & \textbf{Securing Mechanism} & \textbf{Signature Algorithm} & \textbf{Canonicalization} & \textbf{Primary Standard}\tabularnewline
\midrule
SD-JWT VC & JSON claims (JWS Compact + disclosure list) & Enveloping (JWS: JSON Web Signature) & EdDSA (Ed25519) & None & draft-ietf-oauth-sd-jwt-vc-22 \cite{ref1} (selective disclosure: \cite{ref6})\tabularnewline
W3C VCDM & JSON-LD (RDF data model) & Embedded proof (Data Integrity \cite{ref4}, eddsa-rdfc-2022 \cite{ref20}) & Ed25519 & RDFC-1.0 & W3C VCDM 2.0 \cite{ref2}\tabularnewline
mdoc & CBOR & COSE\_Sign1 (MSO: Mobile Security Object) & ECDSA P-256 (ES256) & None & ISO/IEC 18013-5 \cite{ref3}\tabularnewline
\bottomrule
\end{tabular}

\vspace{2pt}
{\footnotesize The digest algorithm is SHA-256 in all three formats: \texttt{\_sd\_alg} in SD-JWT VC, the Hashing Algorithm of Data Integrity, and the MSO digest algorithm in mdoc.}
\end{table*}

\subsection{SD-JWT VC}

\subsubsection{Standards Position}

SD-JWT VC (SD-JWT-based Verifiable Credentials) is a credential format specified as the IETF Internet-Draft draft-ietf-oauth-sd-jwt-vc \cite{ref1}. It adopts JSON Web Token (JWT, RFC 7519 \cite{ref21}) and JSON Web Signature (JWS, RFC 7515) as its foundational technologies, and uses SD-JWT (Selective Disclosure for JSON Web Tokens, RFC 9901 \cite{ref6}) for selective disclosure. It is characterized by high affinity with the existing OAuth 2.0/OpenID Connect ecosystem.

\subsubsection{Data Model and Signing Input}

SD-JWT VC data is represented in JWT Compact Serialization: base64url-encoded Header, Payload and Signature joined by dots. Two properties of that encoding matter here. First, the signing input is the byte string \texttt{header.payload} exactly as transmitted, so no canonicalization step exists and no step can be driven into a worst case. Second, the credential type is carried by the \texttt{vct} claim, and the meaning of claim names is left to the type definition that \texttt{vct} denotes; nothing is resolved from the network while the signature is being checked. The claim set itself (\texttt{iss}, \texttt{sub}, \texttt{vct}, \texttt{iat}, \texttt{exp} and the credential-specific claims) is described in our performance evaluation \cite{ref19}.

\subsubsection{Signature Process}

The signing process for SD-JWT VC follows the JWS specification as follows. First, the Header is constructed and Base64url-encoded. Next, the Payload is constructed and Base64url-encoded. The Signing Input is formed as \texttt{Base64url(Header) || '.' || Base64url(Payload)}, and a signature is generated over this input using the private key. Finally, the signature value is Base64url-encoded, and the compact serialization \texttt{Header.Payload.Signature} is output.

The signature algorithms available to an SD-JWT VC are those registered for JOSE (JWA, RFC 7518, with EdDSA added by RFC 8037)---EdDSA (Ed25519/Ed448), ES256 (ECDSA using P-256 and SHA-256), ES384, ES512 and others; SD-JWT \cite{ref6} itself does not define algorithm identifiers. This experiment used EdDSA (Ed25519).

\subsubsection{Verification Process}

The verification process also follows the JWS specification. The token is split into three parts by dots, and the Header is decoded to obtain the \texttt{alg} header parameter. The verifier checks the alg against an allowlist of acceptable algorithms and rejects algorithms not on the list. This is a defense measure recommended in RFC 8725 \cite{ref22} (JWT Best Current Practices) and serves as a countermeasure against alg:none attacks and algorithm confusion attacks. Next, the signing input is reconstructed, and the signature is verified using the issuer's public key. If the signature is valid, the Payload is decoded and the claims are checked. Of the registered claims the specification lists, only \texttt{vct} is REQUIRED; \texttt{iss}, \texttt{exp}, \texttt{nbf} and \texttt{status} are OPTIONAL, and \texttt{cnf} is REQUIRED only when cryptographic Key Binding is supported.

\subsubsection{Selective Disclosure}

Selective disclosure in SD-JWT VC is realized through the SD-JWT specification \cite{ref6}: disclosable claims in the Payload are replaced by SHA-256 digests and the original values are carried separately as Disclosures. The point that bears on this paper is how the digest is computed. A Disclosure is a JSON array of salt, claim name and claim value, base64url-encoded from its UTF-8 byte sequence, and the digest is taken over that base64url string itself. Because the digest covers the transmitted bytes rather than an abstract structure, no canonicalization is required to absorb differences in whitespace or Unicode escaping---the same property that removes the attack surface studied in Section~\ref{sec:r-poison}.

\subsection{W3C VCDM 2.0 + Data Integrity}

\subsubsection{Standards Position}

VCDM 2.0 + Data Integrity builds on the W3C Verifiable Credentials Data Model 2.0 (VCDM 2.0) \cite{ref2}, which expresses the credential data model in JSON-LD (JSON for Linked Data). It provides semantic interoperability and realizes semantic typing based on the RDF (Resource Description Framework) graph model. Signing uses a cryptosuite that instantiates the W3C Data Integrity specification \cite{ref4}; the suite used here is defined in Data Integrity EdDSA Cryptosuites v1.0 \cite{ref20}. This experiment adopted the eddsa-rdfc-2022 cryptosuite (transformation: RDFC-1.0, hash: SHA-256, signature: EdDSA Ed25519). Ed25519Signature2020 is a legacy Linked Data Proofs cryptosuite, while eddsa-rdfc-2022 is its successor, defined in Data Integrity EdDSA Cryptosuites v1.0 \cite{ref20}; the legacy suite is not evaluated here.

\subsubsection{Data Model}

The data structure of W3C VCDM consists of vocabulary definitions via @context, type, issuer, credentialSubject, and proof (signature information). @context is a core construct of JSON-LD 1.1 \cite{ref23}, containing references to external vocabulary definitions that map JSON keys (property names) to IRIs (Internationalized Resource Identifiers). This ensures that even VCs issued by different issuers have their field semantics unambiguously interpreted by machine, as long as they share the same @context. It is also the one element of the three data models that can send the verifier to the network, and every W3C VCDM attack surface in this paper reaches the verifier through it. A structural example of the serialized credential is given in our performance evaluation \cite{ref19}.

\subsubsection{Data Integrity Proof Specification}

The W3C Verifiable Credential Data Integrity 1.0 specification \cite{ref4} defines a general-purpose framework for generating and verifying cryptographic proofs for VCs. The specification externalizes the selection of specific cryptographic algorithms as cryptosuites (cryptosuite), while the framework itself specifies a three-stage pipeline independent of cryptosuites: Transformation, Hashing, and Proof Generation/Verification.

\textbf{The proof object.} In Data Integrity, signature information is embedded in the VC as a proof object. The key properties of the proof object are those of Table~\ref{tab:2}.

\begin{table}[!t]
\centering
\caption{Table 2. Key Properties of the proof Object (Data Integrity 1.0)}\label{tab:2}
\footnotesize\setlength{\tabcolsep}{4pt}
\begin{tabular}{@{}l@{\hspace{4pt}}l@{\hspace{4pt}}>{\raggedright\arraybackslash}p{0.44\columnwidth}@{}}
\toprule

\textbf{Property} & \textbf{Required} & \textbf{Description}\tabularnewline
type & Required & Proof type (e.g., DataIntegrityProof)\tabularnewline
cryptosuite & Required & Cryptographic suite identifier (e.g., eddsa-rdfc-2022)\tabularnewline
verificationMethod & Optional & Identifier of the public key (verification method) used for verification, expressed as a DID URL or an HTTPS URL. Optional in the specification: when absent, the key is obtained by other means, for instance through \texttt{cryptosuite}. Present in every configuration evaluated here\tabularnewline
proofPurpose & Required & Proof purpose (assertionMethod / authentication, etc.)\tabularnewline
created & Optional & Proof creation datetime (xsd:dateTime)\tabularnewline
domain & Optional & Valid domain for the proof (replay attack prevention)\tabularnewline
challenge & Optional & Challenge value (replay attack prevention)\tabularnewline
proofValue & Required & Signature value (Multibase + Base58btc encoding)\tabularnewline
\bottomrule
\end{tabular}
\end{table}

\subsubsection{Cryptographic Suite Structure}

The Data Integrity specification requires a cryptosuite specification to detail four algorithms. First, the Transformation Algorithm converts the input document to a canonical form suitable for hashing. For eddsa-rdfc-2022, this is RDFC-1.0 canonicalization, which converts the JSON-LD document to an RDF dataset and generates canonical N-Quads through blank node relabeling. Second, the Hashing Algorithm computes a hash of the canonicalized data. SHA-256 is used. Third, the Proof Serialization Algorithm generates the final proof value: Ed25519 signing is performed and the result is Multibase-encoded. Fourth, the Proof Verification Algorithm checks that value against the hash and the verification method.

The cryptosuites the W3C Verifiable Credentials Working Group has published are listed in Table~\ref{tab:3}. All but \texttt{bbs-2023} are Recommendations; \texttt{bbs-2023} is a Candidate Recommendation Draft at the time of writing.

\begin{table*}[!t]
\centering
\caption{Table 3. W3C Data Integrity Cryptographic Suites}\label{tab:3}
\footnotesize\setlength{\tabcolsep}{4pt}
\begin{tabular}{lllll}
\toprule

\textbf{Cryptographic Suite} & \textbf{Transformation} & \textbf{Hash} & \textbf{Sign} & \textbf{Specification}\tabularnewline
eddsa-rdfc-2022 & RDFC-1.0 & SHA-256 & EdDSA (Ed25519) & VC DI EdDSA 1.0\tabularnewline
eddsa-jcs-2022 & JCS (RFC 8785) & SHA-256 & EdDSA (Ed25519) & VC DI EdDSA 1.0\tabularnewline
ecdsa-rdfc-2019 & RDFC-1.0 & SHA-256/384 & ECDSA (P-256/P-384) & VC DI ECDSA 1.0\tabularnewline
ecdsa-jcs-2019 & JCS (RFC 8785) & SHA-256/384 & ECDSA (P-256/P-384) & VC DI ECDSA 1.0\tabularnewline
bbs-2023 & RDFC-1.0 + BBS-specific & SHA-256 & BBS (BLS12-381) & VC DI BBS 1.0\tabularnewline
\bottomrule
\end{tabular}
\end{table*}

The eddsa-rdfc-2022 cryptosuite used in this experiment adopts RDFC-1.0 canonicalization for transformation, SHA-256 for hashing, and EdDSA (Ed25519) for signing. In contrast, eddsa-jcs-2022 adopts JSON Canonicalization Scheme (JCS, RFC 8785) instead of RDFC-1.0, avoiding its computational complexity issues since it does not perform RDF graph canonicalization. However, JCS-based suites do not bind the RDF interpretation of the document to the signature: the document is still JSON-LD and its terms still map to IRIs, but the mapping is outside what the signature covers. The two also differ in what the signature actually protects. With RDFC-1.0 the signing input is the N-Quads obtained after resolving and expanding @context, so the content of the context is incorporated into the signature. With JCS the signing input is the lexical canonical form of the document, in which @context is signed only as a URL string. This difference has security implications, discussed in Section~\ref{sec:disc-vcdm}. Additionally, bbs-2023 is a suite that realizes zero-knowledge proof-based selective disclosure and presentation unlinkability using the BBS signature scheme.

\subsubsection{Signature Process (Add Proof Algorithm)}

The signing process based on the Data Integrity specification's Add Proof algorithm consists of the following steps. We explain using the eddsa-rdfc-2022 cryptosuite as an example.

\textbf{Step 1: Constructing proof options.} Construct a proof options object from metadata such as type, cryptosuite, verificationMethod, proofPurpose, and created. This object does not include the proofValue field (since signing has not yet occurred).

\textbf{Step 2: Document transformation.} Apply the RDFC-1.0 canonicalization algorithm to the input document with the proof property removed. First, resolve the JSON-LD context based on @context and expand the document into an RDF dataset. Then, assign canonical labels to blank nodes in the RDF dataset and output a canonicalized N-Quads format string. This canonicalization process can be the most computationally expensive step, as it determines unique labels for all blank nodes based on graph structure.

\textbf{Step 3: Canonicalization and hashing of proof options.} The proof options object constructed in Step 1 is also treated as a JSON-LD document and similarly undergoes RDFC-1.0 canonicalization. Per \S3.2.5 of \textit{eddsa-rdfc-2022} \cite{ref20}, the \texttt{@context} of the unsecured document is copied into the proof configuration before that canonicalization, so the proof options are expanded under the same context as the document body. A SHA-256 hash is computed over the canonicalized proof options string to obtain the 32-byte proofOptionsHash (proof configuration hash). This step ensures that the proof options content is also protected by the signature.

\textbf{Step 4: Document hashing.} Compute a SHA-256 hash over the canonicalized document string obtained in Step 2 to obtain the 32-byte transformedDocumentHash.

\textbf{Step 5: Hash data concatenation.} Concatenate proofOptionsHash (32 bytes) and transformedDocumentHash (32 bytes) to generate 64-byte hashData. This concatenation ensures that both the proof options and the document body are protected by a single signature.

\textbf{Step 6: Signature generation.} Generate an EdDSA signature over hashData (64 bytes) using the Ed25519 private key. The signature value is stored in the proofValue field as Multibase encoding (Base58btc header \texttt{z} + Base58btc encoding).

\textbf{Step 7: Appending the proof.} Append the proof object containing proofValue to the credential and output the signed VC.

\Needspace*{5\baselineskip}
\textbf{Listing 1. Conceptual Flow of Data Integrity Signing Process}\nopagebreak

\begin{lstlisting}
// Add Proof Algorithm (eddsa-rdfc-2022)
// 1. Construct proof options (excluding proofValue)
proofOptions = {
  type: "DataIntegrityProof",
  cryptosuite: "eddsa-rdfc-2022",
  verificationMethod: "did:example:issuer#key-1",
  proofPurpose: "assertionMethod",
  created: "2024-01-01T00:00:00Z"
}
// 2. Document Transformation
// RDFC-1.0 canonicalization -> N-Quads string
canonicalDoc = RDFC_1_0.canonicalize(
  document_without_proof)
// 3. Canonicalize proof options + hash
canonicalOpts = RDFC_1_0.canonicalize(proofOptions)
proofOptionsHash = SHA256(canonicalOpts) // 32 bytes
// 4. Hash the document
docHash = SHA256(canonicalDoc) // 32 bytes
// 5. Concatenate hash data
hashData = proofOptionsHash || docHash // 64 bytes
// 6. Generate EdDSA signature
signature = Ed25519.sign(hashData, privateKey)
// 7. Multibase-encode as proofValue
proofOptions.proofValue = "z" +
Base58btc.encode(signature)
\end{lstlisting}

An important characteristic of this processing flow is that canonicalization is executed twice: once for the document body and once for the proof options. This ensures that during verification, not only the document content but also the proof metadata (creation date, verification method, purpose, etc.) is covered by the signature. The Sig\_Structure in COSE\_Sign1, which includes the protected header in the signing target, serves a similar role, but W3C VCDM requires explicit canonicalization of both the document and proof options, adding processing overhead.

\subsubsection{Verification Process (Verify Proof in Data Integrity; Proof Verification in the cryptosuite)}\label{sec:diverify}

The Data Integrity specification's Verify Proof algorithm is structured as the inverse operation of Add Proof. The verifier executes the following steps.

\textbf{Step 1: Separating the proof.} Extract the proof object from the VC, obtain and retain the proofValue (signature value), then remove proofValue from the proof object to reconstruct proof options.

\textbf{Step 2: Document transformation and hashing.} Following the same procedure as signing, perform RDFC-1.0 canonicalization on the document excluding proof, and compute the SHA-256 hash (transformedDocumentHash).

\textbf{Step 3: Canonicalization and hashing of proof options.} Perform RDFC-1.0 canonicalization on proof options and compute the SHA-256 hash (proofOptionsHash).

\textbf{Step 4: Hash data concatenation and signature verification.} Reconstruct hashData (64 bytes) by concatenating proofOptionsHash and transformedDocumentHash. Obtain the signature value by Multibase-decoding proofValue, and verify the EdDSA signature using the public key identified by verificationMethod.

An important characteristic of the verification process is the network dependency of canonicalization. HTTP requests are generated to URLs specified in @context, so network latency and context server availability directly affect verification processing time. This characteristic leads to security risks such as SSRF and DoS (see Section~\ref{sec:threat} for the threat model and Sections~\ref{sec:r-tests}, \ref{sec:r-poison} and~\ref{sec:r-loader} for the evaluation results). Using a static document loader (contexts embedded locally) can eliminate network dependency, but requires explicit configuration by application developers.

\subsubsection{RDFC-1.0 Canonicalization Algorithm Details}\label{sec:rdfc}

In the Data Integrity signing and verification pipeline, RDFC-1.0 \cite{ref7}, used in the transformation stage, is an algorithm for canonicalization of RDF datasets. The purpose of canonicalization is to guarantee that two isomorphic RDF datasets have the same string representation. This ensures that for VCs with the same content, the same hash value is always obtained regardless of differences in representation (property order, whitespace, blank node label differences, etc.).

The RDFC-1.0 algorithm consists of three major phases. In the first phase (Initialization), all blank nodes in the input RDF dataset are enumerated, and an initial map computing hash values for each blank node is constructed. In the second phase (Hash First Degree Quads), for each blank node, a hash is computed over the quads in which that blank node appears. In the third phase (Hash N-Degree Quads), for blank nodes not identified in the second phase, recursive hash computation considering a wider range of quads is performed, assigning a unique canonical label to all blank nodes.

The recursive hash computation in the third phase is what causes exponential computational complexity in the worst case. When a cyclic structure (poison graph) where all blank nodes mutually reference each other is input, combinatorial explosion occurs in the N-Degree Quads hash computation. \S4.4.3 of the W3C RDFC-1.0 specification states that implementations MUST defend against potential denial-of-service attacks by raising suitable exceptions and terminating early, and a note there suggests placing a limit (call limit) on the number of Hash N-Degree Quads (\S4.8) invocations based on the number of blank nodes. In the versions we measured, however, no such limit was active by default: in rdf-canonize@3.4.0, which jsonld@8.3.3 uses internally, the call limit \texttt{maxDeepIterations} defaults to \texttt{Infinity}, so no limit applied unless the caller set one explicitly. More importantly, the control is not universally available: no released version of either PyLD (Python) or json-gold (Go) exposes a call limit, as inspection of their sources confirms, so a verifier that installs either from its package registry cannot adopt the mitigation this paper recommends at all. The Rust implementation \texttt{rdf-canon} does provide one, but bounds the total number of Hash N-Degree Quads invocations rather than the per-blank-node count, and ships a default of 4,000. At the time of those measurements the call limit was therefore a parameter the verifier had to set, not one the library supplied. This has since changed on the JavaScript side: rdf-canonize introduced \texttt{maxWorkFactor} in 4.0.0 (November 2023), and in 5.0.0---the version we checked, October 2026, used by jsonld 9.0.0---it has a default of 1 and derives \texttt{maxDeepIterations} from it as \textit{(number of non-unique blank nodes)}\textsuperscript{\texttt{maxWorkFactor}}, so a limit is now in force out of the box. That default is not free of false positives: our replication finds that \texttt{maxWorkFactor}=1 also aborts the acyclic \texttt{forest} inputs at $n=16$ and $20$ and 18 of the RDFC-1.0 specification test-suite cases. The separation this paper reports---every cyclic attack family stopped, every legitimate input admitted---appears at \texttt{maxWorkFactor}=2; at 3, \texttt{ring-16} and \texttt{ring-64} pass again. We notified the maintainers of PyLD and json-gold before publication; the limit introduced in rdf-canonize 4.0.0 was added independently of this work. The recommendation below should therefore be read as two separate questions: whether a limit exists at all, which newer versions answer, and how its threshold is chosen, which remains the verifier's decision because a limit tight enough to stop adversarial topologies can also reject legitimate credentials with many structurally similar blank nodes. Section~\ref{sec:r-poison} measures what the absence of such a limit costs, and Section~\ref{sec:r-limit} sweeps the limit itself to show how tight it has to be.

\subsubsection{Semantic Interoperability}

The greatest technical feature of W3C VCDM is RDF-based semantic interoperability. Since all properties are mapped to globally unique IRIs via @context, the meaning of fields never becomes ambiguous when exchanging credentials between different ecosystems. For example, the field \texttt{issuer} is resolved to the IRI \url{https://www.w3.org/2018/credentials\#issuer} through the W3C VC context and is interpreted as the same meaning in any system. This property plays an essential role in credential interoperability within the DID ecosystem. However, as a trade-off for this semantic interoperability, the aforementioned canonicalization processing incurs performance and security overhead.

\subsection{mdoc (ISO/IEC 18013-5)}

\subsubsection{Standards Position}

mdoc is a credential format for mobile identity documents specified in ISO/IEC 18013-5:2021 \cite{ref3}. It adopts binary encoding via CBOR \cite{ref24} and signing via COSE \cite{ref25}, and is designed with offline presentation via short-range communication (NFC, BLE, Wi-Fi Aware) as its primary use case. ISO/IEC TS 18013-7 specifies extensions for online presentation (OID4VP).

\subsubsection{Data Model}

ISO/IEC 18013-5 defines the mdoc data model in CDDL (Concise Data Definition Language, RFC 8610); the CDDL itself is reproduced in our performance evaluation \cite{ref19}. Two features matter here. Individual data elements are carried as IssuerSignedItem structures whose fields include a \texttt{random} value of at least 16 bytes, which prevents inference of an undisclosed element's content from its digest alone. And the type is given by \texttt{docType} with the claim namespace given structurally by the keys of \texttt{nameSpaces}, so nothing external defines what a claim name means.

The three formats differ markedly in how they convey the credential type and the meaning of claim names. In VCDM 2.0 the type property carries the type, while @context binds claim names to an external vocabulary document. In SD-JWT VC the vct claim carries the type, and the meaning of claim names is left to the type definition it denotes, so no external document is resolved at verification time. mdoc has no counterpart to the external vocabulary document or to a resolvable type definition: \texttt{docType} carries the type and the keys of \texttt{nameSpaces} carry the claim names. Only VCDM 2.0 + Data Integrity therefore resolves an external context during verification, and that difference is what determines the presence or absence of the attack surfaces studied here (context injection and SSRF).

\subsubsection{Mobile Security Object (MSO)}

mdoc signing is performed through the Mobile Security Object (MSO), the structure that aggregates the digests of all data elements and is carried as the payload of IssuerAuth (COSE\_Sign1). Its fields are \texttt{version}, \texttt{digestAlgorithm}, \texttt{valueDigests} (the map of digest IDs to digest values, per namespace), \texttt{deviceKeyInfo}, \texttt{docType} and \texttt{validityInfo}; the full CDDL is in our performance evaluation \cite{ref19}. The security-relevant consequence is that every data element is bound to the signature individually, through its own digest, which is what the tampering tests of Section~\ref{sec:r-tests} exercise.

\subsubsection{Signature Process}

The mdoc signing (issuance) process follows the procedure specified in ISO/IEC 18013-5 \S9.1.2.

\textbf{Stage 1: CBOR encoding of data elements and digest computation.} For each data element (elementIdentifier + elementValue), assign a random salt (random) of at least 16 bytes and a digestID to construct an IssuerSignedItem structure, then CBOR-encode it. For each encoded IssuerSignedItemBytes, individually compute a SHA-256 (or SHA-384/SHA-512) digest. Since this digest is computed over the entire IssuerSignedItemBytes (including the salt), data element contents cannot be inferred from digest values.

\textbf{Stage 2: MSO construction.} Aggregate all data element digests into valueDigests and construct the MobileSecurityObject structure together with digestAlgorithm, deviceKeyInfo, docType, and validityInfo. CBOR-encode the MSO and generate MobileSecurityObjectBytes as a tagged bytestring (CBOR tag 24).

\textbf{Stage 3: COSE\_Sign1 signing.} Based on the COSE\_Sign1 structure \cite{ref25}, generate a signature with MobileSecurityObjectBytes as the payload. The protected header includes the alg parameter (e.g., ES256 = -7), and the Sig\_Structure is constructed. Generate an ECDSA (P-256/P-384/P-521) or EdDSA signature over the Sig\_Structure using the issuer's private key. ISO/IEC 18013-5 specifies four algorithms: ES256, ES384, ES512, and EdDSA. The signer's public key certificate is stored in the COSE unprotected header's x5chain parameter.

\textbf{Stage 4: mdoc assembly.} The COSE\_Sign1 structure (= {[}protected\_header, unprotected\_header, payload, signature{]}) becomes the issuerAuth, and the array of IssuerSignedItemBytes becomes the nameSpaces, assembling the Document structure. The final mdoc is serialized as CBOR binary.

\subsubsection{Verification Process}

mdoc verification (ISO/IEC 18013-5 \S9.3.1 Inspection procedure for issuer data authentication) follows the procedure below.

\textbf{Stage 1: CBOR decoding and structure expansion.} CBOR-decode the received mdoc binary to obtain the Document structure. Expand the issuerAuth (COSE\_Sign1) within issuerSigned to obtain the protected header, payload (MobileSecurityObjectBytes), and signature value.

\textbf{Stage 2: COSE\_Sign1 signature verification.} Obtain the alg parameter from the protected header and confirm it is a permitted algorithm (ES256/ES384/ES512/EdDSA). Reconstruct the same Sig\_Structure as during signing, obtain the issuer's public key certificate from the COSE unprotected header's x5chain, verify the certificate chain (path validation to trust anchor, revocation check), and then verify the signature using the public key. Since the protected header is included in the Sig\_Structure and covered by the signature, protected header tampering is detected as a signature mismatch.

\textbf{Stage 3: Per-element digest verification.} Decode the MSO and obtain the digest map within valueDigests. Compute SHA-256 digests for each received IssuerSignedItemBytes and match them against the corresponding digest values in the MSO (linked by digestID). Data authenticity is confirmed only when all element digests match. This per-element digest method allows the verifier to individually verify the authenticity of only the received elements without needing all data elements.

\subsubsection{Selective Disclosure and Privacy Protection}

Selective disclosure in mdoc follows structurally from the MSO: the holder sends only the issuerSignedItems to be disclosed, and the verifier matches each one's digest against the MSO. Undisclosed elements leave their digests in the MSO, but the \texttt{random} salt makes their contents computationally infeasible to infer. ISO/IEC 18013-5 additionally specifies two forms of mdoc authentication, MAC-based (using an ephemeral key from ECDH, which allows deniable authentication) and signature-based; neither affects the issuer-signature pipeline evaluated here.

\subsubsection{Extension to Online Presentation via ISO/IEC TS 18013-7}

ISO/IEC TS 18013-7 defines an online presentation protocol for mdoc over OID4VP (OpenID for Verifiable Presentations), extending mdoc beyond the face-to-face NFC/BLE presentation of ISO/IEC 18013-5. Because the data model and issuer signature structure are shared, the results of this paper apply to the credential verification step of that protocol as well. The protocol's own security---session encryption, reader authentication, nonce handling and transport-dependent processing---is out of scope here.

\section{Security Evaluation Design}

\subsection{Evaluation Environment}\label{sec:env}

The evaluation in this paper consists of two experiments. The first is a set of security tests in which attack inputs, derived from the threat model defined in Section~\ref{sec:threat}, are submitted to the verification implementation of each format and the resulting behavior---exceptions, rejection, acceptance, and processing time---is observed (Section~\ref{sec:tests}; Sections~\ref{sec:r-tests}--\ref{sec:r-loader}). The second quantifies, as baseline data for estimating denial-of-service exposure, the cost of the canonicalization stage of the verification pipeline by input size and by schema (Section~\ref{sec:threat}; Sections~\ref{sec:r-scaling}--\ref{sec:r-complex}).

The former yields a qualitative verdict on whether an attack succeeds and the latter a quantitative latency value, but this paper defines the success condition for denial of service as a latency increase of 100$\times$ or more over normal processing (Section~\ref{sec:threat}), and it also observes the effectiveness of mitigations as a latency difference; both verdicts therefore presuppose that the observed latency is not buried in the noise of the execution environment. Both experiments were consequently run primarily on a single Linux host reserved exclusively for measurement (Environment~A), and repeated in the three further environments of Table~\ref{tab:4}---a cloud VM with no other workload of ours on it, which is not the same as a physically dedicated host, hence the bare-metal replication described below. Table~\ref{tab:4} gives the details of the evaluation environment. Because attack inputs and cost measurements are handled in the same environment, differences observed between formats cannot be attributed to differences in the execution environment.

\begin{table*}[!t]
\centering
\caption{Table 4. Evaluation Environments. Every experiment in this paper was run in all four. All four environments run the same Node.js build (v24.18.0, V8 13.6, OpenSSL 3.5.7) and the same library versions, so a difference between any two of them is attributable to the hardware. Unless stated otherwise, values quoted in the text are from Environment~A. Common to all four: \texttt{process.hrtime.bigint()} (nanosecond precision); the nominal N is 2,000 after 50 warmup iterations, with the per-condition count set by the kit (at most 200 for the family sweep, 100 for the call-limit sweep, 1,000 for the baseline and the static loader); and five independent process runs per benchmark, of which the median of each statistic is reported.}\label{tab:4}
\footnotesize\setlength{\tabcolsep}{4pt}
\begin{tabular}{@{}l@{\hspace{6pt}}>{\raggedright\arraybackslash}p{0.30\textwidth}@{\hspace{6pt}}>{\raggedright\arraybackslash}p{0.16\textwidth}@{\hspace{6pt}}>{\raggedright\arraybackslash}p{0.17\textwidth}@{\hspace{6pt}}>{\raggedright\arraybackslash}p{0.21\textwidth}@{}}
\toprule
\textbf{Env.} & \textbf{Hardware} & \textbf{OS / kernel} & \textbf{Runtime} & \textbf{Isolation}\tabularnewline
\midrule
A & Linux VM on a public cloud, reserved for measurement (Azure, AMD EPYC 7763, x86\_64); not a physically dedicated host & Ubuntu Linux, 7.0.0-1017-azure & Node.js v24.18.0 & 2 of 4 vCPUs taken offline, SMT disabled; no core pinning, because on this VM pinning widened the run-to-run spread rather than narrowing it \cite{ref19}\tabularnewline
\addlinespace
B & Bare metal, no virtualization (Intel Celeron N5095 @ 2.00\,GHz, x86\_64) & Linux 7.0.0-38-generic & Node.js v24.18.0 & SMT not supported by the CPU, governor \texttt{performance}, turbo disabled (base clock 2.0\,GHz); no core pinning\tabularnewline
\addlinespace
C & Apple M1 Ultra (arm64, 20 cores: 16 performance and 4 efficiency), desktop class, on AC power & macOS 27.0.1 (Darwin 27.0.0) & Node.js v24.18.0 & No core pinning (macOS provides no equivalent); resident backup agents stopped for the measurement\tabularnewline
\addlinespace
D & Intel Xeon E5645 @ 2.40\,GHz (x86\_64, circa 2010, no AVX or AVX2) & Debian 13, Linux 6.12.107 & Node.js v24.18.0 & Pinned with \texttt{numactl} and \texttt{taskset} to one core of one NUMA node; SMT, turbo and frequency scaling disabled\tabularnewline
\bottomrule
\end{tabular}
\end{table*}

The evaluation was performed without any GUI or browser involvement; Environments~A, B and D are server-side Linux and C is a desktop-class macOS machine. Because both the verdict on attack success and the measurement of mitigation effectiveness rely on observing latency, interference from other processes was minimized by disabling SMT and taking unused cores offline; this ensures that observed latency differences are attributable to the attack input and to the configuration of the verification implementation. Environment~D additionally pins the measurement process to a single core; Environments~A, B and C do not---on A because pinning widened the run-to-run spread on that cloud VM rather than narrowing it, and on C because macOS offers no equivalent (Table~\ref{tab:4}). For the canonicalization cost measurements, the suite was executed as five independent process runs and the median of each statistic (mean, $\sigma$, p50, p95, and so on) across runs is reported, mitigating noise that affects any single run; run-to-run variation of p50 was within 3\% for most conditions. Which statistics each environment contributes follows one rule throughout: all four carry p50, because that is the quantity conditions are compared on, together with the ratio against that environment's own baseline where the comparison is a ratio. Table~\ref{tab:10} additionally carries Environment~A's within-run distribution (mean, $\sigma$, p95) as an illustration of its shape; for every other condition that distribution is in the result files rather than in a table. The poison graph sweep and the loader comparison also report the ratio for every environment, because the cross-environment comparison of those ratios is itself a result (Sections~\ref{sec:r-poison} and~\ref{sec:r-loader}). JSON-LD contexts are statically embedded so that no external communication occurs during measurement; the document loader restriction is lifted only when evaluating the attack surface arising from remote context resolution (Section~\ref{sec:r-loader}). Environment A is a cloud VM reserved for measurement rather than a physically dedicated host, so hypervisor-level effects cannot be fully excluded. To compensate, the same kit was run in three further environments: two bare metal without virtualization (B, an Intel Celeron N5095, and D, an Intel Xeon E5645) and one a desktop-class arm64 machine (C, an Apple M1 Ultra). Every experiment was repeated in all of them: the security verdicts (Section~\ref{sec:r-tests}), the poison graph and call-limit sweeps (Sections~\ref{sec:r-poison}--\ref{sec:r-limit}), the loader comparison (Section~\ref{sec:r-loader}) and the canonicalization cost measurements (Sections~\ref{sec:r-scaling}--\ref{sec:r-complex}).

\subsection{Test Tool Architecture}

All experiments in this paper were run with a single measurement kit, vcbench \cite{ref26}: a standalone command-line tool (Node.js, Go, and Python engines plus a shared aggregation script) run without any GUI or web server. Every engine records only raw timings with nanosecond-precision timestamps, and all statistics are computed centrally by the shared script, so the verdicts and the latencies reported below come from the same code path. Four of its suites are used here: \texttt{poison} for the blank-node graph sweep and the call-limit sweep (Sections~\ref{sec:r-poison}--\ref{sec:r-limit}), \texttt{security} for the attack-vector verdicts (Section~\ref{sec:r-tests}), \texttt{loader} for the document loader comparison (Section~\ref{sec:r-loader}), and \texttt{scaling}/\texttt{jsonld-complex} for the canonicalization cost measurements (Sections~\ref{sec:r-scaling}--\ref{sec:r-complex}). The attack-surface suites and the canonicalization cost suites were run separately in each environment, and our performance evaluation \cite{ref19} reports from the same canonicalization runs, so the environment table and every number common to the two papers refer to one dataset. The number of iterations actually completed per condition is recorded in the result files and, where it differs from the nominal $N$, is stated in the table captions. The verification implementations that receive the attack inputs were built with the libraries listed in Table~\ref{tab:5}. The attack surface derives from the specifications, but whether it materializes depends on the libraries and their default settings, so the versions are stated as resolved at measurement time.

\begin{table*}[!t]
\centering
\caption{Table 5. Libraries Used for Each Format (versions resolved from node\_modules at measurement time)}\label{tab:5}
\footnotesize\setlength{\tabcolsep}{4pt}
\begin{tabular}{>{\raggedright\arraybackslash}p{0.121\textwidth}>{\raggedright\arraybackslash}p{0.339\textwidth}>{\raggedright\arraybackslash}p{0.460\textwidth}}
\toprule

\textbf{Format} & \textbf{Primary Library} & \textbf{Role}\tabularnewline
SD-JWT VC & node:crypto (Node.js v24.18.0 / OpenSSL 3.5.7) & EdDSA (Ed25519) signing and verification in the verifier written for this study\tabularnewline
SD-JWT VC & jose@6.2.8 & Reference measurements of a full JWT pipeline only; not used by the verifier under test\tabularnewline
W3C VCDM & jsonld@8.3.3 (rdf-canonize@3.4.0) & JSON-LD processing / RDFC-1.0 canonicalization\tabularnewline
W3C VCDM & node:crypto (Ed25519, SHA-256) & Ed25519 signing / SHA-256 hashing\tabularnewline
W3C VCDM (JCS) & canonicalize@1.0.8 & JCS (RFC 8785) canonicalization\tabularnewline
mdoc & cbor-x@1.6.5 & CBOR encoding and decoding\tabularnewline
mdoc & node:crypto (ECDSA P-256) & ECDSA P-256 signing / COSE\_Sign1 verification\tabularnewline
(Complex credential evaluation) & @digitalcredentials/open-badges-context@3.0.0, @digitalbazaar/credentials-context@3.2.0 & Statically embedded Open Badges v3.0 / VCDM 2.0 contexts (Section~\ref{sec:r-complex})\tabularnewline
\bottomrule
\end{tabular}

\vspace{2pt}
{\footnotesize The jsonld library exposes this canonicalization under the algorithm identifier \texttt{URDNA2015}, which denotes the same algorithm as RDFC-1.0.}
\end{table*}

\subsection{Evaluation Methodology}

\subsubsection{Scope of the Verification Pipeline}\label{sec:scope}

The pipeline evaluated here begins when the verifier accepts an input and ends when the signature over that input has been checked: parsing, resolution and expansion of any context the signing input depends on, canonicalization, hashing, and signature verification. Three steps that a deployed verifier also performs are outside this scope and are not evaluated: retrieval of the issuer's key (DID resolution, issuer metadata), resolution of credential type metadata, and revocation or status checking. Those steps can involve network retrieval in all three formats, so the comparison below concerns the signing pipeline only; statements that one format needs no external reference are to be read within that boundary.

\subsubsection{Threat Model}\label{sec:threat}

Prior to security evaluation, we explicitly state the threat model assumed in this paper. It is consistent with the W3C threat model for the VCDM \cite{ref14} and adds, as evaluation targets, two concerns that the latter does not enumerate as individual threats: the loss of availability caused by the computational cost of verification, and the server-side requests (SSRF) that accompany the resolution of external contexts.

Attacker model: This experiment assumes scenarios where a malicious Holder presents a tampered VP/VC to the verifier, and where a malicious external context provider intervenes in the verifier's processing. The Issuer is assumed to be legitimate. Relative to the Dolev--Yao model that prior VC security work adopts \cite{ref12}, our adversary is weaker in one respect and stronger in another. It is weaker on the network: we do not assume control of the channel between the parties, because the attacks studied here do not need it---the holder is a legitimate participant who simply chooses what to submit. It is stronger on computation: Dolev--Yao idealizes cryptographic operations as atomic and free, which is exactly the abstraction that an algorithmic complexity attack \cite{ref15} breaks, since the adversary's goal is not to forge a signature but to make the verifier spend time. Availability attacks of this kind therefore fall outside what a Dolev--Yao analysis can express, which is one reason they are absent from threat models built on it.

Verifier model: The Verifier is configured as a public API that may accept VCs from arbitrary Issuers. A closed configuration accepting only trusted Issuers is not assumed.

Order of processing: the attacks studied here depend on a property of the Data Integrity verification procedure that we state as an explicit assumption, because without it the assumption of a legitimate Issuer would make the availability argument incoherent. In that procedure (Section~\ref{sec:diverify}) the verifier canonicalizes the document, hashes the result, and only then checks the signature. Canonicalization therefore runs before anything authenticates the input, and an adversary need not hold an issuer key, forge a signature, or even supply a well-formed one: a document with a random byte string in place of the proof value costs the verifier the full canonicalization before it is rejected. Our own measurements bound how lopsided that is. Canonicalizing \texttt{complete-7} costs 1{,}233\,ms (Table~\ref{tab:7}), while the two steps that follow it cost about 0.11\,ms in total---SHA-256 hashing 0.002\,ms and Ed25519 verification 0.107\,ms, both measured with the same kit---so over 99.99\% of the work done to reject an unsigned attack document is done before the signature is examined. The same ordering is what lets the production-schema costs of Section~\ref{sec:r-complex} be imposed at no cost to the sender.

Network assumptions: Default-configured verification implementations are targeted, with W3C VCDM tested with remote context retrieval enabled (no document loader restrictions, no context allowlist configured, no RDFC-1.0 call limit set). One of these omissions is the absence of a control the specification itself defines: \S4.6 of Data Integrity 1.0 \cite{ref4} gives a context validation algorithm that rejects \texttt{@context} values which do not match a known good value or hash, and \S4.1 places it as the fourth step of consuming a secured document. No library we tested applies it by default. The context injection and SSRF results below are therefore measurements of configurations that omit a specified control, not of defects in the formats. This is to detect vulnerabilities in library default configurations. The three formats are, however, not exercised through the same kind of implementation, and the asymmetry must be stated plainly: W3C VCDM is driven through a third-party library stack (\texttt{jsonld}/\texttt{rdf-canonize}) at its own defaults, whereas the SD-JWT VC and mdoc verifiers are written for this study on \texttt{node:crypto} and \texttt{cbor-x} (Table~\ref{tab:5}; \texttt{jose} appears only in the reference pipeline measurement, not in the verifier under test) and follow the practices their specifications recommend---an algorithm allowlist for SD-JWT VC, per-element digest matching against the MSO for mdoc. Results for the latter two therefore characterize what a specification-conformant verifier does, not what an arbitrary third-party implementation does by default. However, the risk mitigation effects of applying production-recommended settings (static document loader, context allowlist, call limit, timeouts, etc.) are discussed in Section~\ref{sec:disc}.

Success criteria: a denial-of-service attack counts as successful when canonicalizing the attack input takes at least 100 times as long as canonicalizing the baseline credential---0.049 ms in Environment A, so the threshold is 4.89 ms. Each condition runs under a 20\,s wall-clock budget, and we record for every one of them which of three outcomes it reached: \emph{completed} (canonicalization returned), \emph{aborted} (the library raised, which under a call limit is the mitigation working), or \emph{truncated} (the iteration count was reduced to stay inside the budget). Reporting the outcome alongside the latency is what keeps an abort from being read as a fast success. For SSRF, the attack surface is confirmed when the processor requests an attacker-named URL; we observe the request at the document loader boundary, so no packet leaves the process (Sections~\ref{sec:r-tests} and~\ref{sec:limits}). Context injection counts as successful when the IRI that a term denotes, or the resulting canonicalized N-Quads, can be made to differ from the issuer's intent.

\subsubsection{Canonicalization Security Tests}\label{sec:tests}

Based on the threat model above, security tests using known attack vectors were conducted for each format. For W3C VCDM, three were tested. Poison graph DoS exploits the worst-case complexity of RDFC-1.0; rather than a single hand-written input we sweep four graph families---the complete graph $K_n$, a bidirectional ring $C_n$, a 3-regular circulant graph, and an acyclic forest as a control---across a range of sizes, and measure the shape of each input after expansion so that the reported node counts are those of the RDF dataset rather than of the generator (Section~\ref{sec:r-poison}). The same families are then re-run across a sweep of the \texttt{maxDeepIterations} call limit, together with four legitimate credentials, to establish whether the limit can be set tightly enough to stop the attacks without rejecting well-formed input (Section~\ref{sec:r-limit}). Context injection substitutes a remotely resolved \texttt{@context} that redefines a term, and we compare the term IRIs and the canonicalized N-Quads before and after. SSRF is tested with a recording document loader: the loader logs the URL the processor asks for and then raises, so the request is observed without being issued. For SD-JWT VC, an \texttt{alg:none} token \cite{ref27} and an algorithm confusion token---an HS256 token whose MAC key is the issuer's Ed25519 public key---were submitted to a verifier with an algorithm allowlist. For mdoc, a modified data element and a modified COSE protected header were submitted to a verifier that checks per-element digests against the MSO. Note that the W3C VCDM tests were conducted with library defaults and no mitigations, which is the configuration in which the attack surfaces exist; Section~\ref{sec:r-limit} measures what one of the mitigations is worth.

\subsubsection{Canonicalization Cost with Production Schemas}\label{sec:costdesign}

The simple credential used as the baseline for the security tests (Section~\ref{sec:tests}) and for the canonicalization cost measurements expands to 5 quads with a single blank node, which never enters the deep recursion, and may therefore underestimate RDFC-1.0 canonicalization cost. To examine whether verification cost can dominate for benign input as well---rather than only for adversarially constructed input---we measured RDFC-1.0 canonicalization time on complex, production-grade schemas (Section~\ref{sec:r-complex}): a 1EdTech Open Badges v3.0 (OB3) AchievementCredential, widely adopted in the education domain, and a sample modeled on the OB3-profile academic degree credentials issued by the Digital Credentials Commons (DCC). OB3 credentials contain multiple nodes without \texttt{id}s (blank nodes in the RDF dataset) under \texttt{achievement}---\texttt{criteria}, \texttt{alignment}, and \texttt{result}---and OB3 is also the representative complex schema used in VC interoperability efforts such as the JFF Plugfest \cite{ref28}, a series of events hosted by Jobs for the Future (JFF) together with the W3C Verifiable Credentials for Education Task Force, in which implementers test their issuers, wallets and verifiers against one another. The JSON-LD contexts (VCDM 2.0 and OB3 v3.0.3) were statically embedded from the \texttt{@digitalcredentials/open-badges-context} and \texttt{@digitalbazaar/credentials-context} packages, eliminating network I/O during measurement. In addition, synthetic credentials containing 10 and 50 id-less child nodes were measured to isolate the effect of blank node count.

The implementation code of the security tests is provided in Appendix A.

\section{Experimental Results}

\subsection{Canonicalization Security Tests}\label{sec:r-tests}

Table~\ref{tab:6} summarises the security test results. Three columns need defining. The \emph{severity} column rates, on a qualitative three-level scale of our own (we do not use CVSS or any other established scoring system), the magnitude of the impact if the attack succeeds under default settings with no mitigation applied. CRITICAL denotes an attack that bypasses signature verification or reaches the verifier environment or its internal network, and that requires almost no special precondition to succeed. HIGH denotes an attack that compromises the integrity of claims or the availability of the verification process, but whose success requires an additional condition such as a particular implementation setting, key configuration or input scale. NONE denotes a case in which the attack surface does not exist at all, by construction of the format. Severity is assigned to the potential impact of the attack surface and is deliberately independent of what this experiment detected.

What the experiment detected is recorded in the other two columns, which we separate because they answer different questions. The \emph{evidence} column states how the row is established: \emph{measured} means the behavior was produced by submitting the input to a running verifier in this study, with the outcome recorded by the kit; \emph{by construction} means the format has no such processing stage, so the claim follows from the specification and nothing is measured. The \emph{outcome} column states what happened. The separation matters because an attack that a verifier we ran actually rejected and an attack surface that cannot exist are different findings, and a single verdict column conflates them.

All nine rows are either measured or true by construction; none rests on inspection of source code alone. Seven rows are measured: the six verdicts of the \texttt{security} suite, discussed in this subsection, and the poison graph row, which the \texttt{poison} suite establishes and Section~\ref{sec:r-poison} reports. Of those seven, four are rejections and three are successful attacks. The SD-JWT VC verifier rejected the unsigned \texttt{alg:none} token (\textit{alg not allowed: none}) and a forged HMAC token whose MAC key was the issuer's Ed25519 public key (\textit{alg not allowed: HS256}); the mdoc verifier rejected both a modified data element (\textit{digest mismatch}) and a modified COSE protected header (\textit{COSE\_Sign1 signature verification failed}). On the W3C VCDM side, context injection changed the IRI that a term denotes---\texttt{http://schema.org/name} became \url{http://attacker.example.com/vocab#displayName}, and the canonicalized N-Quads differed accordingly---and the SSRF test observed the processor asking its document loader for \url{http://169.254.169.254/latest/meta-data/}, the cloud instance metadata endpoint. The loader in that test records the URL and refuses it rather than issuing the request, so reachability is established by observing the request at the loader boundary instead of by putting a packet on the wire; Section~\ref{sec:limits} states why we drew the boundary there. Poison graph denial of service, the third successful attack, exceeds the 100$\times$ criterion by a wide margin and is the subject of Section~\ref{sec:r-poison}. The six verdicts of the \texttt{security} suite, and the detail strings reproduced above, are identical in all four environments and across all five runs in each. This is what one would expect of outcomes that depend on the specification rather than on timing, and reporting it is the check that they do.

\begin{table*}[!t]
\centering
\caption{Table 6. Security Test Results Summary (all outcomes produced by the \texttt{security} suite of vcbench, except the poison graph row, which comes from the \texttt{poison} suite; the verdicts and their detail strings are identical in all four environments and across all five runs in each)}\label{tab:6}
\footnotesize\setlength{\tabcolsep}{4pt}
\begin{tabular}{@{}>{\raggedright\arraybackslash}p{0.20\textwidth}@{\hspace{4pt}}>{\raggedright\arraybackslash}p{0.105\textwidth}@{\hspace{4pt}}l@{\hspace{4pt}}>{\raggedright\arraybackslash}p{0.165\textwidth}@{\hspace{4pt}}>{\raggedright\arraybackslash}p{0.29\textwidth}@{}}
\toprule

\textbf{Test item} & \textbf{Target} & \textbf{Severity} & \textbf{Evidence} & \textbf{Outcome}\tabularnewline
\midrule
Poison graph DoS (RDFC-1.0) & W3C VCDM & HIGH & Measured (Section~\ref{sec:r-poison}) & \textbf{Attack succeeds}: 278$\times$ baseline from 5 blank nodes in 527\,B (251--424$\times$ across the four environments)\tabularnewline
Context injection via \texttt{@context} & W3C VCDM & HIGH & Measured (Section~\ref{sec:r-tests}) & \textbf{Expansion redirected}: the term IRI changes when the context is resolved remotely. Under \texttt{eddsa-rdfc-2022} verification then fails, so the measured impact is availability; the integrity impact is argued for the JCS and enveloping paths (\S7)\tabularnewline
SSRF via remote context & W3C VCDM & CRITICAL & Measured at the loader boundary (Section~\ref{sec:r-tests}, Section~\ref{sec:disc}) & \textbf{Attack surface confirmed}: loader asked for 169.254.169.254\tabularnewline
No canonicalization & SD-JWT VC / mdoc & NONE & By construction & Stage absent\tabularnewline
\texttt{alg:none} & SD-JWT VC & CRITICAL & Measured (Section~\ref{sec:r-tests}) & Rejected: \textit{alg not allowed: none}\tabularnewline
Algorithm confusion (EdDSA$\rightarrow$HS256) & SD-JWT VC & HIGH & Measured (Section~\ref{sec:r-tests}) & Rejected: \textit{alg not allowed: HS256}\tabularnewline
Data element tampering & mdoc & HIGH & Measured (Section~\ref{sec:r-tests}) & Rejected: \textit{digest mismatch}\tabularnewline
COSE protected header tampering & mdoc & HIGH & Measured (Section~\ref{sec:r-tests}) & Rejected: \textit{COSE\_Sign1 signature verification failed}\tabularnewline
No external reference in the signing pipeline & SD-JWT VC / mdoc & NONE & By construction & Stage absent\tabularnewline
\bottomrule
\end{tabular}
\end{table*}

\subsection{Poison Graph Denial of Service}\label{sec:r-poison}\label{sec:poison}

Table~\ref{tab:7} reports the cost of RDFC-1.0 canonicalization for adversarially constructed blank node graphs under the library defaults, with no call limit applied. By \emph{poison graph} we mean a JSON-LD document whose nodes carry explicit blank node identifiers (\texttt{\_:b}$i$) and reference one another, so that RDFC-1.0 cannot separate them by their immediate surroundings and must enter the Hash N-Degree Quads recursion to assign canonical labels \cite{ref7,ref8}. We sweep four families parameterized by a node count $n$: the complete graph $K_n$, in which every blank node references every other; a bidirectional ring $C_n$; a 3-regular circulant graph (\emph{cubic}); and, as a control, the acyclic forest produced by a generator that omits \texttt{@id}. The shape of each input is measured after expansion rather than assumed---quad count, blank node count, whether some blank node appears both as the subject and as the object of a reference (a necessary condition for a cycle, and the test the harness applies), and document size---because the number of blank nodes in the resulting RDF dataset is not in general the generator's parameter. Each condition runs under a 20\,s wall-clock budget, and conditions whose iteration count was reduced to stay within it are noted in the caption. The baseline of 0.049\,ms used here is the same simple credential as the 0.047\,ms of Table~\ref{tab:11}; the two figures come from different suites (\texttt{poison} and \texttt{jsonld-complex}) with different iteration counts, measured minutes apart in the same session on the same kernel. The 4\% gap between them is comparable to the run-to-run variation of Section~\ref{sec:env}; we report each figure against its own suite rather than reconciling them. Each table is read against the baseline measured alongside it. The full sweep was run in all four environments, so the table reports each one against its own baseline.

\begin{table*}[!t]
\centering
\caption{Table 7. RDFC-1.0 Canonicalization of Adversarial Blank Node Graphs in All Four Environments (\texttt{jsonld} 8.3.3 defaults, no call limit; p50 of the medians across 5 runs). N=200 per condition except where the 20\,s wall-clock budget reduced it; the budget bites at different sizes on different machines, so the reduced conditions differ by environment and the count actually completed is recorded per run in the result files. Mean, $\sigma$ and p95 for every condition are in the result files. Ratios are against each environment's own baseline (A: 0.049\,ms, B: 0.122\,ms, C: 0.022\,ms, D: 0.168\,ms), so the 100$\times$ success criterion of Section~\ref{sec:threat} corresponds to 4.89, 12.25, 2.18 and 16.79\,ms respectively; these are 100 times the unrounded baselines, which is why they do not reproduce exactly from the three-digit values shown. Bold marks the smallest input of each family that crosses the criterion in that environment: it is the same input in all four for the three cyclic families, and not for the acyclic control.}\label{tab:7}
\scriptsize\setlength{\tabcolsep}{2pt}
\begin{tabular}{@{}l l r r r c r r r r r r r r@{}}
\toprule
& & & & & & \multicolumn{2}{c}{\textbf{Env.~A}} & \multicolumn{2}{c}{\textbf{Env.~B}} & \multicolumn{2}{c}{\textbf{Env.~C}} & \multicolumn{2}{c}{\textbf{Env.~D}}\tabularnewline
\cmidrule(lr){7-8}\cmidrule(lr){9-10}\cmidrule(lr){11-12}\cmidrule(lr){13-14}
\textbf{Input} & \textbf{Family} & \textbf{Quads} & \textbf{BN} & \textbf{Bytes} & \textbf{Cyc.} & \textbf{p50} & \textbf{$\times$base} & \textbf{p50} & \textbf{$\times$base} & \textbf{p50} & \textbf{$\times$base} & \textbf{p50} & \textbf{$\times$base}\tabularnewline
\midrule
Simple credential & baseline & 5 & 1 & 612 & no & 0.049 & 1.0$\times$ & 0.122 & 1.0$\times$ & 0.022 & 1.0$\times$ & 0.168 & 1.0$\times$\tabularnewline
\texttt{complete-3} & $K_n$ & 6 & 3 & 231 & yes & 0.302 & 6.2$\times$ & 0.897 & 7.3$\times$ & 0.170 & 7.8$\times$ & 0.993 & 5.9$\times$\tabularnewline
\texttt{complete-4} & $K_n$ & 12 & 4 & 364 & yes & 1.774 & 36$\times$ & 4.829 & 39$\times$ & 1.253 & 58$\times$ & 6.020 & 36$\times$\tabularnewline
\texttt{complete-5} & $K_n$ & 20 & 5 & 527 & yes & 13.576 & \textbf{278$\times$} & 36.946 & \textbf{302$\times$} & 9.232 & \textbf{424$\times$} & 42.127 & \textbf{251$\times$}\tabularnewline
\texttt{complete-6} & $K_n$ & 30 & 6 & 720 & yes & 123.584 & 2{,}529$\times$ & 336.154 & 2{,}745$\times$ & 80.179 & 3{,}683$\times$ & 388.618 & 2{,}314$\times$\tabularnewline
\texttt{complete-7} & $K_n$ & 42 & 7 & 943 & yes & 1{,}233.259 & 25{,}235$\times$ & 3{,}366.976 & 27{,}491$\times$ & 788.800 & 36{,}233$\times$ & 3{,}874.658 & 23{,}072$\times$\tabularnewline
\texttt{complete-8} & $K_n$ & 56 & 8 & 1{,}196 & yes & 13{,}494.092 & 276{,}117$\times$ & 36{,}454.022 & 297{,}643$\times$ & 8{,}594.325 & 394{,}769$\times$ & 42{,}180.759 & 251{,}174$\times$\tabularnewline
\texttt{cubic-6} & circulant & 18 & 6 & 540 & yes & 5.278 & \textbf{108$\times$} & 15.499 & \textbf{127$\times$} & 3.597 & \textbf{165$\times$} & 17.980 & \textbf{107$\times$}\tabularnewline
\texttt{cubic-8} & circulant & 24 & 8 & 716 & yes & 11.726 & 240$\times$ & 30.218 & 247$\times$ & 7.412 & 340$\times$ & 36.065 & 215$\times$\tabularnewline
\texttt{cubic-10} & circulant & 30 & 10 & 892 & yes & 25.781 & 528$\times$ & 69.168 & 565$\times$ & 16.199 & 744$\times$ & 81.261 & 484$\times$\tabularnewline
\texttt{cubic-16} & circulant & 48 & 16 & 1{,}444 & yes & 201.892 & 4{,}131$\times$ & 538.751 & 4{,}399$\times$ & 125.837 & 5{,}780$\times$ & 649.374 & 3{,}867$\times$\tabularnewline
\texttt{ring-8} & $C_n$ & 16 & 8 & 596 & yes & 1.747 & 36$\times$ & 4.778 & 39$\times$ & 1.058 & 49$\times$ & 5.903 & 35$\times$\tabularnewline
\texttt{ring-16} & $C_n$ & 32 & 16 & 1{,}198 & yes & 7.634 & \textbf{156$\times$} & 20.576 & \textbf{168$\times$} & 4.660 & \textbf{214$\times$} & 25.379 & \textbf{151$\times$}\tabularnewline
\texttt{ring-64} & $C_n$ & 128 & 64 & 4{,}846 & yes & 188.885 & 3{,}865$\times$ & 483.947 & 3{,}951$\times$ & 123.190 & 5{,}659$\times$ & 563.373 & 3{,}355$\times$\tabularnewline
\texttt{ring-128} & $C_n$ & 256 & 128 & 9{,}794 & yes & 1{,}058.804 & 21{,}665$\times$ & 2{,}695.516 & 22{,}009$\times$ & 741.911 & 34{,}079$\times$ & 3{,}024.411 & 18{,}009$\times$\tabularnewline
\texttt{forest-20} & acyclic & 80 & 60 & 3{,}112 & no & 0.966 & 20$\times$ & 2.661 & 22$\times$ & 0.537 & 25$\times$ & 3.002 & 18$\times$\tabularnewline
\texttt{forest-100} & acyclic & 400 & 300 & 15{,}592 & no & 4.700 & 96$\times$ & 12.736 & \textbf{104$\times$} & 2.815 & \textbf{129$\times$} & 14.478 & 86$\times$\tabularnewline
\texttt{forest-500} & acyclic & 2{,}000 & 1{,}500 & 78{,}792 & no & 26.287 & \textbf{538$\times$} & 70.311 & 574$\times$ & 15.551 & 714$\times$ & 77.625 & \textbf{462$\times$}\tabularnewline
\bottomrule
\end{tabular}
\end{table*}

Three of the four families cross the 100$\times$ criterion with inputs smaller than 1.2\,kB---\texttt{complete-5}, at 527 bytes, is smaller than the 612-byte baseline credential, yet costs 278$\times$ as much. They do so in all four environments---at the same input in each, since the four run identical runtime versions. \texttt{complete-5}---5 blank nodes in a 527-byte document---takes 13.576\,ms (p50) in Environment~A, or 278$\times$ the baseline, and 302$\times$, 424$\times$ and 251$\times$ in B, C and D; \texttt{cubic-6} (6 blank nodes, 540\,B) is at 108$\times$, 127$\times$, 165$\times$ and 107$\times$; \texttt{ring-16} (16 blank nodes, 1{,}198\,B) at 156$\times$, 168$\times$, 214$\times$ and 151$\times$. Above that, the complete family grows by roughly an order of magnitude for each blank node added: in Environment~A, 124\,ms at 6 nodes, 1.23\,s at 7, and 13.5\,s at 8; the slowest environment costs 389\,ms, 3.87\,s and 42.2\,s for the same inputs and the fastest 80\,ms, 789\,ms and 8.59\,s. The last of these comes from a 1{,}196-byte document---an amplification between $2.5 \times 10^{5}$ and $3.9 \times 10^{5}$ over the baseline depending on the environment, from an input that would pass any plausible request size limit. The denial-of-service exposure documented in the RDFC-1.0 security considerations \cite{ref8} is thus reproducible at the implementation level on four different machines, and the cost is governed by the structure of the input rather than by its size.

The four environments also show that the multiple is not the invariant here, and they show why. The baseline and the attack inputs do not scale together: relative to Environment~A the baseline is 2.51$\times$ slower in B, 3.44$\times$ slower in D and 0.45$\times$ as slow in C, while the attack inputs move by 2.55--2.97$\times$, 2.86--3.41$\times$ and 0.56--0.71$\times$ respectively. Which of the two moves further is not the same in every direction, and this is what keeps the multiple from being a transferable quantity. Going from A to C the baseline speeds up by more than the attacks do, so the multiple rises; going from A to B the attacks slow down by slightly more than the baseline does, so the multiple rises there as well; going from A to D the two move together and it falls slightly. The result is that the multiple does not order the environments by speed: for \texttt{complete-5} it is C 424$\times$, B 302$\times$, A 278$\times$, D 251$\times$, whereas by baseline speed the machines run C, A, B, D. All four share the same Node.js build and library versions, so this is a hardware effect and not a runtime-version artifact, and the same ordering appears in every family. Two consequences follow for the criterion. First, the smallest crossing input of each cyclic family is the same in all four environments---\texttt{complete-5}, \texttt{cubic-6} and \texttt{ring-16}---while the acyclic forest, the control that should not cross at all, crosses at \texttt{forest-100} in Environments~B and C (104$\times$ and 129$\times$) and only at \texttt{forest-500} in A and D. Second, and more plainly: a criterion phrased as a multiple of normal processing is met at different absolute costs on different machines, and in a way that does not follow from how fast the machine is, so a verifier cannot infer safety either from a slow machine or from the control family's margin.

The acyclic forest family is the control, and it is the more instructive result. At $n=20$ it contains 60 blank nodes, nearly four times as many as \texttt{ring-16}, yet reaches only 18--25$\times$ across the four environments (0.966\,ms in Environment~A); scaled to 1{,}500 blank nodes in a 79\,kB document it reaches 462--714$\times$ (26.3\,ms in Environment~A), a cost that grows roughly linearly in input size rather than explosively. Because its nodes carry no explicit identifiers, the generator yields a forest of disjoint three-node components: two id-less nodes pointing at one shared blank node. Their blank nodes are not distinguishable by their first-degree hashes, so RDFC-1.0 does call Hash N-Degree Quads for every one of them; each call sees one component and recurses a single level, so the cost stays linear in the number of components. What drives RDFC-1.0 into its worst case is therefore the symmetry of the mutual references among blank nodes, not their number---a distinction with direct consequences for mitigation, since a bound on blank node count alone would admit \texttt{complete-8} while rejecting the harmless \texttt{forest-500}.

\subsection{DoS Mitigation via the RDFC-1.0 Call Limit}\label{sec:r-limit}

The \texttt{rdf-canonize} backend of the \texttt{jsonld} library accepts \texttt{maxDeepIterations}, an upper bound $k$ on the number of iterations of the Hash N-Degree Quads recursion; when the bound is exceeded, canonicalization aborts with an exception instead of running to completion. In the version measured here (\texttt{rdf-canonize} 3.4.0 via \texttt{jsonld} 8.3.3) it defaults to \texttt{Infinity} (Section~\ref{sec:rdfc}), and Table~\ref{tab:7}---where canonicalization ran for 13.5\,s without aborting---is the direct evidence that the recursion is unbounded unless the caller sets the option. This control is distinct from a processing timeout in two ways. A timeout bounds elapsed time regardless of why it was consumed, whereas a call limit bounds the specific recursion that the attack exploits and so can be set from a property of the application's schemas rather than from a latency budget. More consequentially, a timeout imposed from outside the canonicalizer does not stop the work. Our replication measured this directly across three runtimes: a 2{,}000\,ms timeout wrapped around the call (\texttt{Promise.race} on Node.js, a goroutine with \texttt{select} in Go, a thread in Python) returns to the caller at about 2{,}000\,ms, but canonicalization continues to occupy a core for a further 0.5 to 31 seconds. On Node.js it does not even fire for the ring inputs, because the canonicalizer yields to the event loop only during its first 15\,ms. Only controls that interrupt the computation itself stop it: a SIGALRM delivered to PyLD's main thread, and the in-algorithm time limit of the Java implementation titanium-rdfc. The \texttt{AbortSignal} added in \texttt{rdf-canonize} 4.0.0, which we checked in 5.0.0, is checked once every three permutations, which is too rarely to help on rings---$n=512$ ran to completion in 48 seconds. For the purposes of this paper, therefore, ``timeout'' means a control that halts the computation, and a wall-clock timeout wrapped around the call is not one. Table~\ref{tab:8} sweeps $k$ over $\{1, 4, 16, 64\}$ for both the attack families of Section~\ref{sec:r-poison} and a set of legitimate credentials, so that the limit's selectivity---whether it rejects attacks without rejecting well-formed input---can be read off directly.

\begin{table*}[!t]
\centering
\caption{Table 8. Effect of the RDFC-1.0 Call Limit (\texttt{maxDeepIterations}=$k$) on Legitimate and Adversarial Input, in All Four Environments (p50 in ms, medians across 5 runs, N=100 per cell; the single exception is \texttt{ring-64} at $k$=64 in Environment~D, where the time budget reduced it to 68--69. $\dagger$ = canonicalization aborted with ``Maximum deep iterations ($k$) exceeded'', so the time shown is the cost of detecting the attack, not of canonicalizing it). Which cells abort is identical in all four environments.}\label{tab:8}
\footnotesize\setlength{\tabcolsep}{4pt}
\begin{tabular}{@{}l l r c r r r r@{}}
\toprule
\textbf{Input} & \textbf{Kind} & \textbf{BN} & \textbf{Env.} & \textbf{$k$=1} & \textbf{$k$=4} & \textbf{$k$=16} & \textbf{$k$=64}\tabularnewline
\midrule
Simple credential & legitimate & 1 & A & 0.049 & 0.042 & 0.041 & 0.041\tabularnewline
 &  &  & B & 0.158 & 0.121 & 0.117 & 0.119\tabularnewline
 &  &  & C & 0.028 & 0.023 & 0.021 & 0.021\tabularnewline
 &  &  & D & 0.174 & 0.156 & 0.150 & 0.149\tabularnewline
\addlinespace
Synthetic, 10 id-less nodes & legitimate & 10 & A & 0.189 & 0.180 & 0.174 & 0.168\tabularnewline
 &  &  & B & 0.561 & 0.514 & 0.537 & 0.477\tabularnewline
 &  &  & C & 0.107 & 0.099 & 0.098 & 0.094\tabularnewline
 &  &  & D & 0.581 & 0.551 & 0.521 & 0.504\tabularnewline
\addlinespace
Synthetic, 50 id-less nodes & legitimate & 50 & A & 0.742 & 0.744 & 0.741 & 0.742\tabularnewline
 &  &  & B & 1.892 & 1.860 & 1.865 & 1.862\tabularnewline
 &  &  & C & 0.443 & 0.435 & 0.438 & 0.437\tabularnewline
 &  &  & D & 2.013 & 1.987 & 1.989 & 1.988\tabularnewline
\addlinespace
Open Badges v3.0 credential & legitimate & 4 & A & 1.290 & 1.258 & 1.253 & 1.239\tabularnewline
 &  &  & B & 3.152 & 3.077 & 3.061 & 3.039\tabularnewline
 &  &  & C & 0.863 & 0.876 & 0.865 & 0.857\tabularnewline
 &  &  & D & 3.352 & 3.248 & 3.229 & 3.198\tabularnewline
\addlinespace
\texttt{complete-5} & attack & 5 & A & 0.389$\dagger$ & 0.687$\dagger$ & 1.914$\dagger$ & 6.962$\dagger$\tabularnewline
 &  &  & B & 1.041$\dagger$ & 1.841$\dagger$ & 5.203$\dagger$ & 18.929$\dagger$\tabularnewline
 &  &  & C & 0.229$\dagger$ & 0.420$\dagger$ & 1.231$\dagger$ & 4.724$\dagger$\tabularnewline
 &  &  & D & 1.140$\dagger$ & 2.106$\dagger$ & 6.127$\dagger$ & 22.741$\dagger$\tabularnewline
\addlinespace
\texttt{complete-7} & attack & 7 & A & 0.700$\dagger$ & 1.392$\dagger$ & 4.155$\dagger$ & 15.310$\dagger$\tabularnewline
 &  &  & B & 1.818$\dagger$ & 3.693$\dagger$ & 11.150$\dagger$ & 40.638$\dagger$\tabularnewline
 &  &  & C & 0.422$\dagger$ & 0.871$\dagger$ & 2.626$\dagger$ & 9.909$\dagger$\tabularnewline
 &  &  & D & 2.067$\dagger$ & 4.294$\dagger$ & 13.197$\dagger$ & 49.049$\dagger$\tabularnewline
\addlinespace
\texttt{cubic-10} & attack & 10 & A & 0.360$\dagger$ & 0.733$\dagger$ & 1.893$\dagger$ & 8.350$\dagger$\tabularnewline
 &  &  & B & 0.959$\dagger$ & 1.921$\dagger$ & 5.021$\dagger$ & 22.335$\dagger$\tabularnewline
 &  &  & C & 0.208$\dagger$ & 0.424$\dagger$ & 1.158$\dagger$ & 5.306$\dagger$\tabularnewline
 &  &  & D & 1.077$\dagger$ & 2.238$\dagger$ & 6.003$\dagger$ & 27.221$\dagger$\tabularnewline
\addlinespace
\texttt{ring-16} & attack & 16 & A & 0.677$\dagger$ & 1.385$\dagger$ & 4.161$\dagger$ & 7.639\tabularnewline
 &  &  & B & 1.789$\dagger$ & 3.667$\dagger$ & 11.172$\dagger$ & 20.448\tabularnewline
 &  &  & C & 0.392$\dagger$ & 0.828$\dagger$ & 2.461$\dagger$ & 4.583\tabularnewline
 &  &  & D & 2.076$\dagger$ & 4.379$\dagger$ & 13.555$\dagger$ & 25.086\tabularnewline
\addlinespace
\texttt{ring-64} & attack & 64 & A & 3.572$\dagger$ & 7.901$\dagger$ & 26.256$\dagger$ & 96.656$\dagger$\tabularnewline
 &  &  & B & 9.327$\dagger$ & 20.441$\dagger$ & 67.225$\dagger$ & 246.145$\dagger$\tabularnewline
 &  &  & C & 2.290$\dagger$ & 5.124$\dagger$ & 16.994$\dagger$ & 62.732$\dagger$\tabularnewline
 &  &  & D & 10.503$\dagger$ & 23.685$\dagger$ & 77.490$\dagger$ & 287.341$\dagger$\tabularnewline
\addlinespace
\texttt{forest-20} & attack (acyclic) & 60 & A & 0.964 & 0.964 & 0.967 & 0.967\tabularnewline
 &  &  & B & 2.617 & 2.618 & 2.612 & 2.609\tabularnewline
 &  &  & C & 0.520 & 0.521 & 0.539 & 0.540\tabularnewline
 &  &  & D & 3.024 & 3.021 & 3.014 & 3.015\tabularnewline
\bottomrule
\end{tabular}
\end{table*}

Three observations follow. First, the limit is selective: every legitimate input completes at every $k$, including $k=1$---the simple credential (0.049\,ms), synthetic credentials with 10 and 50 id-less child nodes (0.189 and 0.742\,ms) and an Open Badges v3.0 credential (1.290\,ms). None of these requires the deep recursion at all, so the bound never binds. The p50 does drift down from $k$=1 to $k$=64---for the simple credential by 16, 25, 25 and 14\% in Environments~A to D---but it does so in the same direction in all four and only for the first column measured, which is the signature of warm-up rather than of the bound. Second, every cyclic attack family is stopped at $k \le 16$, aborting with ``Maximum deep iterations ($k$) exceeded''. The single exception appears at $k=64$, where \texttt{ring-16} completes in 7.639\,ms (156$\times$ the baseline): a limit of 64 is already too loose for that family, which places the usable range for the inputs tested here at $k \le 16$. Third, the cost of the abort itself grows with $k$---\texttt{ring-64} costs 3.572\,ms to reject at $k=1$ but 26.256\,ms at $k=16$---so $k$ should be set as low as the application's legitimate schemas permit, and a verifier that accepts only credentials of known shape can defensibly set it to a single-digit value.

The same three observations hold in Environments~B, C and D, with the same boundaries. Every legitimate input completes at every $k$; every cyclic attack family aborts at $k \le 16$; \texttt{ring-16} is in each case the single condition that completes at $k=64$ (20.448, 4.583 and 25.086\,ms); and \texttt{forest-20} completes at every $k$. Which cells abort is identical in all four environments---the same nineteen of the forty, with the same error---although the time each abort takes is not, since that is a latency. That the usable range for the limit comes out identical across machines spanning a factor of eight in speed is the useful part: unlike a latency budget, a call limit counts work the input causes rather than time the machine takes, so a threshold chosen on one deployment carries to another. The separate question of whether the \emph{ratio} by which an input exceeds the baseline is stable across machines is taken up below, and the answer there is different.

The limit is nevertheless necessary rather than sufficient, for the reason established in Section~\ref{sec:r-poison}: it bounds the depth of the recursion per blank node, so an input whose recursion is shallow but wide passes whatever the bound. \texttt{forest-20} completes unchanged at every $k$, and the 26.3\,ms that \texttt{forest-500} costs is likewise untouched by any call limit. A call limit must therefore be combined with a bound on document size or blank node count, and with a control that actually halts the computation when it overruns---an in-algorithm time check, or a timeout enforced where the work can be killed, such as a separate process. As the replication cited above records, a timeout wrapped around the canonicalization call returns to the caller without stopping it, so a verifier that relies on one keeps paying for the attack after it has answered the request. The three controls cover different inputs and none of them subsumes the others.

Two caveats attach to recommending this control. The first is availability: of the implementations we are aware of, \texttt{rdf-canonize} (JavaScript) and \texttt{rdf-canon} (Rust) expose a limit, while no released PyLD or json-gold does, so on those stacks the recommendation reduces to a document size bound and a timeout. Where a limit does exist its unit differs---\texttt{maxDeepIterations} bounds the recursion per blank node, whereas \texttt{rdf-canon} bounds the total invocation count with a default of 4,000, which admits a 9-blank-node torus---a family from that replication, not one of ours---costing some 2{,}500$\times$ the baseline---so a threshold cannot be carried across implementations unchanged. Of the implementations we have seen, only \texttt{rdf-canon} (Rust) and titanium-rdfc (Java) are architecturally independent of jsonld.js; PyLD, json-gold and \texttt{rdf-canonize} 5.0.0 all descend from it, so agreement among those three is weaker evidence than their number suggests. The second is that our 100$\times$ criterion is a ratio, and inputs can sit close to it. \texttt{ring-16} clears it in all four of our environments (156$\times$, 168$\times$, 214$\times$ and 151$\times$), and so does \texttt{cubic-6}, though only just in two of them (108$\times$, 127$\times$, 165$\times$ and 107$\times$). The input whose verdict actually changes with the machine is in the control family: \texttt{forest-100} crosses in Environments~B and C (104$\times$ and 129$\times$) and not in A or D (96$\times$ and 86$\times$). The families and the shape of their growth are robust; for an input that sits near the line, the ratio criterion's verdict is a property of the machine as much as of the input, and the smallest $n$ at which a given family crosses the threshold should not be read as a constant of the algorithm.

\subsection{JSON-LD Context Loader Comparison}\label{sec:r-loader}

The security of W3C VCDM canonicalization depends on how \texttt{@context} is resolved, and so does its latency. Table~\ref{tab:9} compares three document loader configurations on a single document: a minimal credential with one context and two properties, smaller than the simple credential of Table~\ref{tab:7}, so the 0.023\,ms static figure is not comparable with the 0.047--0.049\,ms baseline used elsewhere. The comparison is between loaders, not between documents. The \emph{static} loader resolves every context from a map embedded in the measurement process, so no request is issued at all. The \emph{delayed} loader is the static document loader plus an injected delay; the delay is reported in the table so that the resulting ratio is not mistaken for a measured network cost. The \emph{loopback} loader performs a real HTTP retrieval from a server that the measurement process starts itself, which gives the cost of actual retrieval with the network distance reduced to zero.

\begin{table*}[!t]
\centering
\caption{Table 9. JSON-LD Document Loader Comparison in All Four Environments (medians across 5 runs; N=1,000 for the static document loader and N=100 for the other two; ratios are against each environment's own static loader. The injected delay of the delayed condition is a parameter of the experiment, not a measured network latency.)}\label{tab:9}
\footnotesize\setlength{\tabcolsep}{3pt}
\begin{tabular}{@{}l l r r r r r r r r@{}}
\toprule
& & \multicolumn{2}{c}{\textbf{Env.~A}} & \multicolumn{2}{c}{\textbf{Env.~B}} & \multicolumn{2}{c}{\textbf{Env.~C}} & \multicolumn{2}{c}{\textbf{Env.~D}}\tabularnewline
\cmidrule(lr){3-4}\cmidrule(lr){5-6}\cmidrule(lr){7-8}\cmidrule(lr){9-10}
\textbf{Loader} & \textbf{Retrieval} & \textbf{p50} & \textbf{$\times$static} & \textbf{p50} & \textbf{$\times$static} & \textbf{p50} & \textbf{$\times$static} & \textbf{p50} & \textbf{$\times$static}\tabularnewline
\midrule
Static (contexts embedded) & none & 0.023 & 1.0$\times$ & 0.062 & 1.0$\times$ & 0.011 & 1.0$\times$ & 0.087 & 1.0$\times$\tabularnewline
Loopback (real HTTP, 0 hops) & HTTP to \texttt{127.0.0.1} & 0.473 & \textbf{20.4$\times$} & 1.178 & \textbf{18.9$\times$} & 0.275 & \textbf{25.9$\times$} & 0.967 & \textbf{11.2$\times$}\tabularnewline
Delayed (+50\,ms injected) & none (delay simulated) & 50.231 & 2{,}166$\times$ & 50.702 & 812$\times$ & 52.223 & 4{,}915$\times$ & 50.353 & 581$\times$\tabularnewline
\bottomrule
\end{tabular}
\end{table*}

The loopback condition is the informative one, because it isolates the cost of retrieval from the cost of the network. Even with client and server in the same host and no hop between them, resolving a context over HTTP costs 0.473\,ms (p50) against 0.023\,ms for a locally embedded context---20.4$\times$---and the tail widens with it, p95 rising from 0.060\,ms to 0.597\,ms in the result files. That figure is a lower bound: any real deployment adds its own round-trip time to the 0.473\,ms, so the penalty for remote resolution starts at a factor of 20 and is thereafter determined by the network rather than by the verifier. The other three environments give factors of 18.9, 25.9 and 11.2 (B, C and D). As with the poison ratios, these do not line up with machine speed: the retrieval cost and the static baseline do not scale together, and the fastest machine shows the largest factor while the slowest shows the smallest. What the four agree on is the floor, which is an order of magnitude everywhere. The delayed condition is reported for completeness and should be read as what it is: with 50\,ms injected, p50 becomes 50.231\,ms, and the resulting 2{,}166$\times$ ratio restates the injected delay divided by the static cost rather than measuring any property of the library. We therefore quote the loopback figure, not the delayed one, when comparing loader configurations.

The security difference is categorical rather than quantitative. A loader that retrieves contexts is the configuration in which both the context injection and the SSRF results of Section~\ref{sec:r-tests} exist at all: the processor requests whatever URL the credential names, which is how the recording loader observed the request for \url{http://169.254.169.254/latest/meta-data/}, and it expands whatever document comes back, which is how the term IRI \texttt{http://schema.org/name} became \url{http://attacker.example.com/vocab#displayName}. Embedding contexts statically removes both at once, because there is no request to redirect and no document to substitute. The latency reduction is a secondary benefit of a change made for security reasons.

\subsection{Attribute-Count Scaling of Canonicalization Cost}\label{sec:r-scaling}

\begin{figure}[!t]
\centering
\includegraphics[width=\columnwidth]{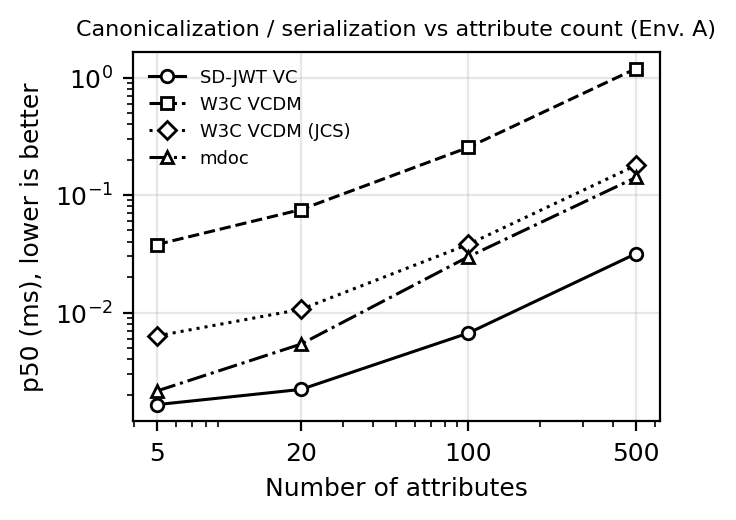}
\caption{Figure 1. Attribute-Count Scaling of Canonicalization and Serialization Cost. Horizontal axis: number of attributes in the credential (5, 20, 100, 500, logarithmic). Vertical axis: p50 time for one serialization in milliseconds, logarithmic; lower is better. One line per format; W3C VCDM includes RDFC-1.0 canonicalization, SD-JWT VC and mdoc do not. Data as in Table~\ref{tab:10}.}\label{fig:sec1}
\end{figure}

How canonicalization cost grows with input size is baseline data for estimating denial-of-service exposure. Table~\ref{tab:10} and Fig.~\ref{fig:sec1} show the change in serialization speed (without cryptographic processing; W3C VCDM includes RDFC-1.0 canonicalization via the \texttt{jsonld} library, whereas for SD-JWT VC and mdoc the measured step is encoding, which a verifier does not perform) when the number of attributes is varied over 5, 20, 100, and 500. Measurements used \texttt{process.hrtime.bigint()} at N=2,000 (N=400 for W3C VCDM at 100 attributes and above) $\times$ 5 runs.

\begin{table*}[!t]
\centering
\caption{Table 10. Attribute Count Scaling (serialization speed vs.\ attribute count; medians across 5 runs; N=2,000 per condition except $\dagger$=400. Mean, $\sigma$, p95 and ops/sec are Environment~A; p50 is given for all four environments.)}\label{tab:10}
\footnotesize\setlength{\tabcolsep}{3pt}
\begin{tabular}{@{}l r r r r r r r r r r@{}}
\toprule
& & \multicolumn{5}{c}{\textbf{Env.~A}} & \textbf{B} & \textbf{C} & \textbf{D} & \tabularnewline
\cmidrule(lr){3-7}\cmidrule(lr){8-10}
\textbf{Format} & \textbf{Attrs} & \textbf{mean} & \textbf{ops/s} & \textbf{$\sigma$} & \textbf{p50} & \textbf{p95} & \textbf{p50} & \textbf{p50} & \textbf{p50} & \textbf{Size (bytes)}\tabularnewline
\midrule
SD-JWT VC & 5 & 0.003 & 608,643 & 0.014 & 0.002 & 0.004 & 0.004 & 0.001 & 0.005 & 272\tabularnewline
W3C VCDM & 5 & 0.053 & 26,413 & 0.133 & 0.038 & 0.073 & 0.112 & 0.020 & 0.142 & 681\tabularnewline
W3C VCDM (JCS) & 5 & 0.007 & 158,428 & 0.008 & 0.006 & 0.008 & 0.014 & 0.004 & 0.017 & 581\tabularnewline
mdoc & 5 & 0.003 & 466,418 & 0.015 & 0.002 & 0.004 & 0.005 & 0.001 & 0.005 & 355\tabularnewline
\addlinespace
SD-JWT VC & 20 & 0.003 & 451,671 & 0.004 & 0.002 & 0.004 & 0.005 & 0.001 & 0.006 & 732\tabularnewline
W3C VCDM & 20 & 0.084 & 13,304 & 0.058 & 0.075 & 0.104 & 0.196 & 0.041 & 0.241 & 1,716\tabularnewline
W3C VCDM (JCS) & 20 & 0.011 & 93,985 & 0.006 & 0.011 & 0.011 & 0.023 & 0.006 & 0.028 & 926\tabularnewline
mdoc & 20 & 0.006 & 185,874 & 0.002 & 0.005 & 0.008 & 0.014 & 0.004 & 0.015 & 1,300\tabularnewline
\addlinespace
SD-JWT VC & 100 & 0.007 & 149,633 & 0.007 & 0.007 & 0.010 & 0.016 & 0.004 & 0.020 & 3,185\tabularnewline
W3C VCDM $\dagger$ & 100 & 0.278 & 3,930 & 0.061 & 0.254 & 0.427 & 0.645 & 0.143 & 0.744 & 7,236\tabularnewline
W3C VCDM (JCS) & 100 & 0.039 & 26,281 & 0.007 & 0.038 & 0.041 & 0.074 & 0.024 & 0.078 & 2,766\tabularnewline
mdoc & 100 & 0.030 & 33,506 & 0.005 & 0.030 & 0.033 & 0.075 & 0.020 & 0.070 & 6,417\tabularnewline
\addlinespace
SD-JWT VC & 500 & 0.035 & 31,617 & 0.031 & 0.032 & 0.046 & 0.082 & 0.022 & 0.083 & 15,452\tabularnewline
W3C VCDM $\dagger$ & 500 & 1.245 & 840 & 0.151 & 1.191 & 1.446 & 2.986 & 0.663 & 2.844 & 34,836\tabularnewline
W3C VCDM (JCS) & 500 & 0.182 & 5,598 & 0.012 & 0.179 & 0.191 & 0.337 & 0.117 & 0.343 & 11,966\tabularnewline
mdoc & 500 & 0.138 & 7,091 & 0.028 & 0.141 & 0.155 & 0.356 & 0.096 & 0.329 & 32,262\tabularnewline
\bottomrule
\end{tabular}
\end{table*}

W3C VCDM (RDFC-1.0 canonicalization) serialization time rose from 0.038 ms (p50) at 5 attributes to 1.191 ms at 500 attributes (a 31$\times$ increase for a 100$\times$ increase in attributes; 37$\times$ SD-JWT VC at 500 attributes). The growth rate itself rises with scale: the per-interval exponent (the slope on log-log axes) increases monotonically from 0.49 for 5--20 attributes to 0.76 for 20--100 and 0.96 for 100--500, so the cost approaches linear growth at the top of the range tested. All four environments reproduce the same shape, with the exponent rising monotonically over the same three intervals: 0.40/0.74/0.95 in B, 0.52/0.78/0.95 in C and 0.38/0.70/0.83 in D.

How this growth translates into denial-of-service resilience is clearest in terms of throughput. On a single core, canonicalization alone sustains roughly 21,200 credentials/s for a simple credential (0.047 ms (p50)), about 840/s at 500 attributes (1/25 of that), and about 620/s for the production OB3 schema (Section~\ref{sec:r-complex}; 1.603 ms (p50)), i.e. 1/34. A verifier whose capacity was sized on simple credentials therefore loses about 97\% of it merely by accepting OB3-shaped traffic; from an attacker's point of view, the same input amplifies the verifier's work roughly 34-fold at essentially no cost to the sender.

None of the inputs measured here reaches our success criterion for denial of service (4.89\,ms of canonicalization, Section~\ref{sec:threat}): at 500 attributes the cost is 1.191 ms, and extrapolating the measured exponent of 0.96 puts the threshold at roughly 2,200 attributes. The comparison with Section~\ref{sec:r-poison} is the point of reporting this. Growing a credential along the attribute axis is a poor route to denial of service---the growth is close to linear, so the attacker pays in bandwidth roughly what the verifier pays in time---whereas the adversarial topologies of Section~\ref{sec:r-poison} cross the same threshold with five blank nodes in 527 bytes. The attribute axis is the one benign traffic moves along, and the blank node axis is the one an attacker moves along; they are different threats and the cost curves say so. By comparison, the formats without canonicalization scale almost flatly over the same range: SD-JWT VC from 0.002 to 0.032 ms, JCS from 0.006 to 0.179 ms, and mdoc from 0.002 to 0.141 ms (all p50). A fuller treatment of these cross-format scaling differences, including payload sizes, is given in our performance evaluation \cite{ref19}; here we carry forward only what bears on availability.

\subsection{Canonicalization Cost with Production Schemas}\label{sec:r-complex}

Denial of service does not require adversarial input to become a concern: the canonicalization cost of schemas already deployed in production can by itself dominate verification. Table~\ref{tab:11} contrasts the simple credential used as the baseline (5 quads, one blank node) with complex credentials based on production schemas. The evaluated credentials are: a 1EdTech Open Badges v3.0 (OB3) AchievementCredential \cite{ref29}, widely used in the education domain (containing id-less nodes such as \texttt{criteria}, \texttt{alignment}, and \texttt{result}); a sample modeled on the OB3-profile academic degree credentials issued by the Digital Credentials Commons (DCC); and synthetic credentials with 10 and 50 id-less child nodes to isolate the effect of blank node count. All JSON-LD contexts are statically embedded; no network I/O is included in the measurements.

\begin{table*}[!t]
\centering
\caption{Table 11. RDFC-1.0 Canonicalization of Complex Credentials in All Four Environments (\texttt{jsonld} 8.3.3 over \texttt{rdf-canonize} 3.4.0; p50 in ms, medians across 5 runs; N=1,000, $\dagger$N=200. \emph{vs simple} is the ratio against each environment's own simple credential.)}\label{tab:11}
\footnotesize\setlength{\tabcolsep}{3pt}
\begin{tabular}{@{}l r r r r r r r r r r@{}}
\toprule
& & & \multicolumn{2}{c}{\textbf{Env.~A}} & \multicolumn{2}{c}{\textbf{Env.~B}} & \multicolumn{2}{c}{\textbf{Env.~C}} & \multicolumn{2}{c}{\textbf{Env.~D}}\tabularnewline
\cmidrule(lr){4-5}\cmidrule(lr){6-7}\cmidrule(lr){8-9}\cmidrule(lr){10-11}
\textbf{Credential} & \textbf{Quads} & \textbf{BN} & \textbf{p50} & \textbf{vs simple} & \textbf{p50} & \textbf{vs simple} & \textbf{p50} & \textbf{vs simple} & \textbf{p50} & \textbf{vs simple}\tabularnewline
\midrule
Simple credential & 5 & 1 & 0.047 & 1.0$\times$ & 0.122 & 1.0$\times$ & 0.021 & 1.0$\times$ & 0.170 & 1.0$\times$\tabularnewline
Open Badges v3.0 $\dagger$ & 32 & 4 & 1.603 & \textbf{34.1$\times$} & 3.924 & \textbf{32.1$\times$} & 1.039 & \textbf{49.5$\times$} & 5.298 & \textbf{31.1$\times$}\tabularnewline
DCC-style degree credential $\dagger$ & 21 & 2 & 1.092 & 23.2$\times$ & 2.650 & 21.7$\times$ & 0.709 & 33.8$\times$ & 2.772 & 16.3$\times$\tabularnewline
Synthetic (10 blank nodes) & 43 & 10 & 0.187 & 4.0$\times$ & 0.501 & 4.1$\times$ & 0.098 & 4.7$\times$ & 0.526 & 3.1$\times$\tabularnewline
Synthetic (50 blank nodes) $\dagger$ & 203 & 50 & 0.748 & 15.9$\times$ & 1.864 & 15.3$\times$ & 0.428 & 20.4$\times$ & 1.991 & 11.7$\times$\tabularnewline
\bottomrule
\end{tabular}
\end{table*}

\begin{figure}[!t]
\centering
\includegraphics[width=\columnwidth]{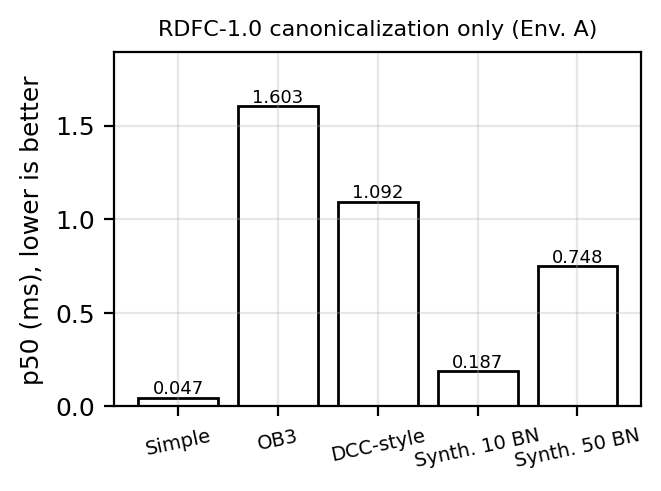}
\caption{Figure 2. RDFC-1.0 Canonicalization of Complex Credentials. Horizontal axis: credential, ordered as in Table~\ref{tab:11}. Vertical axis: p50 canonicalization time in milliseconds; lower is better. All contexts statically embedded, so no network retrieval is included. Data as in Table~\ref{tab:11}.}\label{fig:sec2}
\end{figure}

Canonicalizing the OB3 credential took 1.603 ms (p50) in Environment~A---roughly 34$\times$ the simple credential (0.047 ms). The DCC-style credential took about 23$\times$ (1.092 ms (p50)). Fig.~\ref{fig:sec2} plots the five credentials on a common axis. The multiple depends on the environment: it is 34.1$\times$, 32.1$\times$, 49.5$\times$ and 31.1$\times$ for the OB3 credential in A, B, C and D, and 23.2$\times$, 21.7$\times$, 33.8$\times$ and 16.3$\times$ for the DCC-style credential, with the ordering of the five credentials identical in all four. All four run the same Node.js build and library versions (Table~\ref{tab:4}), so the spread among them is a hardware effect. Canonicalizing OB3 dwarfs the rest of the verification in every environment, so the conclusion that canonicalization dominates the verification pipeline for production schemas holds in all four environments.

Two factors drive the increase, and separating them matters for mitigation. The first is expansion of OB3's large JSON-LD context (hundreds of term definitions); the second is blank node canonical labeling. The synthetic credentials isolate the second: 10 id-less nodes cost 0.187 ms and 50 cost 0.748 ms, a factor of 4.0 for five times the nodes, i.e. an exponent of 0.86 over that interval---steep, but far below the cost of OB3. OB3 itself points the same way: with only 4 blank nodes it costs 1.603 ms, more than twice the synthetic credential with 50. The two inputs differ in quad count as well (32 against 203), so this does not isolate expansion from labeling; separating them would need the expansion step timed on its own. What it does show is that blank node labeling is not the dominant term for these production schemas, and a guard expressed as an upper bound on blank node count would not address it. Pre-compiled contexts and expansion caching are the controls that apply. For OB3 and DCC-style credentials the mean-median gap and $\sigma$ are also large (OB3 in Environment~A: mean 2.067 ms with $\sigma$ 1.688, from the result files), indicating that complex context processing has a long latency tail.

From a security standpoint, this means that the conditions for a canonicalization-induced denial of service are not confined to adversarially constructed poison graphs. A verifier that accepts credentials from arbitrary issuers can be driven into canonicalization costs one to two orders of magnitude above the baseline by ordinary, well-formed credentials, and the cost grows steeply with blank node count as well as with attribute count. In domains where W3C VCDM adoption is advancing with education credentials (OB3/DCC), mitigations such as pre-compiled contexts, expansion caching, and a bound on document size are therefore not merely performance optimizations but availability controls. SD-JWT VC and mdoc require no canonicalization, so schema complexity has limited impact on their verification cost.

\section{Discussion}\label{sec:disc}

\subsection{Security of VCDM 2.0 + Data Integrity}\label{sec:disc-vcdm}

The JSON-LD/RDFC verification pipeline has additional attack surfaces derived from remote context resolution and RDFC-1.0 canonicalization. This paper's experiments were conducted with default settings (no mitigations), and all three vulnerability types (poison graph DoS, context injection, SSRF) can be mitigated with appropriate production settings: static document loader, context allowlist, RDFC-1.0 call limit, and HTTP request timeouts. The first two of these are not ad hoc hardening: they are the two implementation strategies that \S4.6 of Data Integrity 1.0 \cite{ref4} names as equivalent to its context validation algorithm. The same two are also recommended independently of the Verifiable Credentials specifications: the JSON-LD Working Group's Best Practices draft \cite{ref30} asks that systems processing JSON-LD choose an explicit policy for resolving contexts, lists context spoofing and processing delay among the risks of resolving them over the network, and recommends that implementations provision vetted local representations of the contexts they expect and resolve only those. The call limit has no counterpart there---that draft addresses how a context is resolved, not the cost of canonicalizing what was resolved---so it is the one of the three for which no guidance we are aware of states a threshold, which is why Section~\ref{sec:r-limit} measures one. However, these mitigations require explicit configuration by developers, and the risk of vulnerability exposure due to default settings can potentially manifest.

Regarding poison graph DoS, the RDFC-1.0 algorithm requires worst-case exponential complexity in the canonical labeling of blank nodes \cite{ref8,ref13}, and Section~\ref{sec:r-poison} shows that the worst case is cheap to construct: 5 blank nodes in 527 bytes cost 278$\times$ the baseline, and 8 in 1,196 bytes cost 13.5 seconds, with both results reproduced on separate hardware. Two consequences follow for practice. The first is that the attack is not gated by input size, so the request size limits a verifier already has provide no protection at all---an adversary needs a kilobyte, not a megabyte. The second is more specific, and it is the result we would most want implementers to take away: cost is driven by the symmetry of the references among blank nodes, not by their number. Our acyclic control carries 1,500 blank nodes in 79 kB and still costs only 538$\times$, roughly an eighth of \texttt{cubic-16} with its 16 nodes. A guard expressed purely as an upper bound on blank node count would therefore reject harmless documents while admitting the dangerous ones; the specification's own note \cite{ref8} points the other way when it suggests bounding the number of Hash N-Degree Quads invocations, which is the call limit of Section~\ref{sec:r-limit}, and should be replaced or supplemented by a call limit on the recursion itself (Section~\ref{sec:r-limit}), which keys on the behavior rather than on a proxy for it.

The availability concern is nevertheless not confined to adversarial input, and this is a separate finding with a separate mitigation. Canonicalizing a simple credential took 0.047 ms (p50), whereas a benign production schema (OB3) took 1.603 ms (about 34$\times$, Table~\ref{tab:11}) and a synthetic credential with 50 blank nodes took 0.748 ms (about 16$\times$); the same trend was reproduced in Environments~B to D (about 32$\times$, 49$\times$ and 31$\times$ for OB3). A call limit does nothing about this, because legitimate schemas never enter the recursion it bounds (Table~\ref{tab:8}). Capacity planning, context precompilation and expansion caching are the controls that apply here, and a verifier needs both sets: one against constructed input, one against the traffic it expects.

Regarding context injection, it is worth stating the conditions under which the attack succeeds. With eddsa-rdfc-2022 the signature is computed over the N-Quads obtained after expanding @context, so rewriting the @context of a signed document changes the canonicalization result and the signature no longer verifies. The concern addressed here is therefore the configuration in which the verifier resolves a remote context under the attacker's control instead of a locally trusted one: the expansion result then depends on what was resolved, and the IRI that a term denotes can differ from the issuer's intent. Our test confirmed that terms can be overridden, and canonicalization still succeeds, when the context lacks the @protected keyword of JSON-LD 1.1 \cite{ref23}. A verifier that performs the context validation of \S4.6 of Data Integrity 1.0 \cite{ref4} would not reach this point at all, since the substituted context matches no known value; what the test characterizes is the configuration in which that step is skipped. This is why the outcome column of Table~\ref{tab:6} qualifies the row---the attack succeeds \emph{where the context is resolved remotely and lacks} \texttt{@protected}; it is not a demonstration that arbitrary reinterpretation is possible while the signature remains valid.

The protection against this attack therefore depends on the securing mechanism. On the RDFC-1.0 path just described, a change of meaning is detected cryptographically, because the expansion is itself the signing input. With a cryptosuite based on JCS (eddsa-jcs-2022), or with enveloping protection in which the credential is wrapped in JOSE/COSE, the signing input contains only the URL string of the context. Substituting the document that the URL resolves to then leaves the signature valid while changing the IRI a term denotes---that is, the meaning of the claims---and because verification succeeds, the verifier is left with no signal of the inconsistency. This last statement follows from what each cryptosuite defines as its signing input, not from an experiment: the substitution test reported in Section~\ref{sec:r-tests} was run on the RDFC-1.0 path, where it is the canonicalization result that changes. Constructing the corresponding worked example on the JCS and enveloping paths---signing a credential, substituting the document its context URL resolves to, and showing that verification still succeeds---would turn the argument into a demonstration, and we leave it to future work.

This also affects how integrity protection of external resources should be regarded. The W3C threat model \cite{ref14} lists, as a mitigation for T7, a mechanism that binds a digest of each related resource into the credential, and leaves its adoption to implementers. On paths other than RDFC-1.0, the signature itself cannot bind the meaning of a context, so the protection has to come from elsewhere: either integrity protection of related resources, or restricting contexts to known-good local copies---the static loader of Section~\ref{sec:r-loader}. One of the two is in effect a requirement rather than an option. W3C VCDM (JCS), which this paper also measures, avoids the computational complexity of RDFC-1.0 but comes with this reduction in protection scope.

A further mitigation follows from the ordering stated in Section~\ref{sec:threat} rather than from the context mechanism itself: because canonicalization precedes authentication, any check that can reject an input more cheaply should be moved ahead of it. Resolving \texttt{verificationMethod} against an issuer allowlist, or rejecting documents whose \texttt{proof} is structurally malformed, costs microseconds and disqualifies an attacker who holds no issuer key---which, under the threat model of this paper, is every attacker. This does not help against a legitimate issuer's own heavy schema, which is why it complements rather than replaces the controls above. Combining a context allowlist with @protected, and using statically embedded contexts, are effective defenses. The latter follows the same approach as the mitigation that the W3C threat model \cite{ref14} lists for T7---resolving external resources from a local store of known-good values rather than fetching them at processing time---and it removes a latency penalty as well: Section~\ref{sec:r-loader} measures retrieval over a loopback interface, with no network between the parties, at 20.4$\times$ the cost of a statically embedded context, which is the floor that any real deployment builds on.

Regarding SSRF, a JSON-LD processor may send HTTP requests to URLs specified in @context. If an attacker specifies a cloud metadata endpoint (http://169.254.169.254/) or an internal network address, the structure permits the verifier to issue requests to those addresses. Section~\ref{sec:r-tests} observes this directly: given a credential whose \texttt{@context} names \url{http://169.254.169.254/latest/meta-data/}, the processor asked its document loader for exactly that URL. The loader used in the test records the request and raises instead of performing it, so what we establish is reachability---the verifier will ask for whatever the credential names---rather than a completed retrieval from a live metadata service. We stop at that boundary deliberately (Section~\ref{sec:limits}): everything after it is a property of the network the verifier sits on, not of the verification pipeline, and the pipeline is what this paper evaluates. The CRITICAL severity in Table~\ref{tab:6} reflects the impact should that boundary be the only thing standing in the way, which in a default configuration it is. Not only @context but any external resource referenced from a credential lies outside the protection of the securing mechanism (T7 of the W3C threat model \cite{ref14}), and fetching it is itself observable as a request issued by the verifier. As a countermeasure, an allowlist must be applied in the document loader so that only trusted context URLs are permitted. Our tool embeds contexts in advance and thereby eliminates network retrieval altogether.

\subsection{Security of SD-JWT VC}\label{sec:disc-sdjwt}

In tests against SD-JWT VC, both the alg:none token and the algorithm confusion token were rejected. It is worth being precise about what rejected them, because the answer is not a library. The verifier used here is written for this study on \texttt{node:crypto} and checks the \texttt{alg} header against an explicit allowlist of \texttt{EdDSA} before it does anything else; both tokens fail on that check, reporting \textit{alg not allowed: none} and \textit{alg not allowed: HS256} respectively, and neither reaches a signature operation. What the result demonstrates is therefore that the practice RFC 8725 (JWT Best Current Practices) \cite{ref22} prescribes---fixing the acceptable algorithms at the verifier rather than reading them from the token---is sufficient against both attacks, not that any particular library implements it. The distinction matters because the allowlist is precisely what a verifier can forget to configure: libraries have shipped that accept alg:none \cite{ref27}, and in those that do not, the protection is usually a parameter the caller supplies. An implementer should read this row of Table~\ref{tab:6} as a property of the configuration, not of the dependency.

\subsection{Security of mdoc}

mdoc correctly detected tampering in both tampering detection tests verified in this paper (data element tampering, COSE header tampering). Each data element is individually SHA-256 digested, and the digest is included in the MSO (signed by COSE\_Sign1), so even single-byte modifications are detected as digest mismatches. COSE protected header tampering is also detected because the protected header is included in the Sig\_Structure and covered by the signature. These results confirm the robustness of mdoc's data integrity protection mechanism within the scope of the evaluated attack vectors.

\section{Limitations and Future Work}\label{sec:limits}

This study has several limitations. First, the measurement environment has SMT disabled and unused cores offline, but it is a cloud VM reserved for measurement rather than a physically dedicated host, so hypervisor-level effects cannot be fully excluded; core pinning, which would be the usual remedy, widened the run-to-run spread on this VM rather than narrowing it \cite{ref19}, so the session reported here is unpinned. To compensate, every experiment was repeated in three further environments, two of them bare metal without virtualization (Environment~B, Intel Celeron N5095, and Environment~D, Intel Xeon E5645) and one a desktop-class arm64 machine (Environment~C, Apple M1 Ultra). The absolute values differ and so do the ratios---by up to 1.7$\times$ between the largest and smallest, and not in the order of machine speed (Section~\ref{sec:r-poison})---but every qualitative finding is reproduced: the same families cross the criterion with sub-kilobyte inputs, the same call limit range separates attack from legitimate input, and the same six security verdicts are returned. We further mitigate run-to-run noise by reporting medians of statistics across five independent runs, reporting outlier counts in the result files, and interpreting results primarily through p50 (run-to-run p50 variation was within 3\% for most conditions). JIT optimization and garbage collection effects nevertheless remain; stricter measurement with systems languages such as C or Rust is desirable.

Second, the poison graphs we construct come from four hand-chosen families, selected because their symmetry is easy to reason about. They establish that the worst case is reachable with small inputs, but they are not a search: families outside this set may be cheaper still per byte, and resource-guided fuzzing \cite{ref17} would be the way to look for them. Relatedly, the sweep stops where a single canonicalization exceeds our 20\,s budget, so the exponent is established over the range we could measure rather than extrapolated.

Third, the SSRF test observes the request at the document loader boundary---the loader records the URL the processor asks for and refuses it---rather than allowing the request onto the network. This is a deliberate scope decision rather than a shortcoming of the method: whether a request to 169.254.169.254 returns credentials depends on the cloud provider, the instance metadata service version and the host's network policy, none of which is a property of the verification pipeline. What a sandbox experiment would add is a demonstration that a particular deployment is exploitable; what it would not change is the finding that the verifier issues the request.

Fourth, the figures in our tables come from a single implementation stack (\texttt{jsonld} 8.3.3 over \texttt{rdf-canonize} 3.4.0 on Node.js). Our replication across PyLD, json-gold, \texttt{rdf-canonize} 5.0.0, the Rust \texttt{rdf-canon} and the Java titanium-rdfc reproduces the shape of the growth and the ordering of the families in all five, so the phenomenon is not an artifact of one library. What does remain library-specific is the constants, the availability of a call limit at all---no released PyLD or json-gold offers one, and titanium-rdfc bounds elapsed time rather than invocations---and the unit and default of the limit where one exists.

Future work includes constructing the worked example for the JCS and enveloping paths discussed in Section~\ref{sec:disc-vcdm}, evaluating how the policy adopted when an external resource such as a status list cannot be retrieved (fail-open versus fail-closed) affects availability, searching for worst-case graph families by resource-guided fuzzing, quantifying the risk reduction achieved by production-equivalent configurations with mitigations applied, measuring the families against implementations we have not yet covered,  and analyzing the attack surface of additional signature suites such as BBS+.

\textit{Ethical considerations.} The behavior this paper measures is not an undisclosed vulnerability. RDFC-1.0 documents dataset poisoning in \S7.1 and requires implementations to defend against it in \S4.4.3; what we add is the price of the missing default, in bytes and in milliseconds on commodity hardware. We nevertheless notified the maintainers of PyLD and json-gold before publication, those being the two stacks in our survey where no released version offers the control we recommend, and offered the limit sweep of Section~\ref{sec:r-limit} so that a default can be chosen against measurements rather than intuition. We did not request an embargo: an embargo on behavior the specification already describes would delay mitigation guidance without withholding anything an attacker could not derive from the specification itself. The poison generators and the measurement harness are published in \texttt{vcbench} for the same reason---a verifier cannot establish whether its own stack aborts without an input that makes it abort. We did not approach the maintainers of \texttt{rdf-canonize}, \texttt{rdf-canon} or titanium-rdfc, each of which already ships a limit, and we measured only systems we operate ourselves.

\section{Threats to Validity}

We discuss threats to validity that may affect the generalizability of our results.

\textit{Internal validity.} The canonicalization cost measurements were performed with \texttt{process.hrtime.bigint()} on the Node.js V8 runtime, so JIT optimization stages, GC pauses, and event-loop interference may affect the results. We report medians of statistics across five independent runs, each preceded by 50 warmup iterations, with the measured iteration count per condition as given in Table~\ref{tab:4} and each caption (100 to 2,000, and as low as 1 where a single canonicalization exceeded the 20\,s budget); outliers (Tukey 1.5$\times$IQR, the interquartile range) are reported without removal, and p50---robust to outliers---is used as the representative statistic. GC effects nonetheless cannot be entirely eliminated. Measurements were performed with SMT disabled and without core pinning (Table~\ref{tab:4}) on a host reserved for measurement, but that host is a cloud VM rather than a physically dedicated machine, so hypervisor-level effects cannot be fully excluded; replication in three further environments, two of them bare metal (B and D), confirms that the canonicalization-cost trend is reproduced (Section~\ref{sec:r-complex}), although hardware dependence remains and the multiples themselves vary with the environment. Environment~C has no core pinning, because macOS provides no equivalent, and is therefore the least tightly isolated of the four; we report it as such.

\textit{External validity.} The schemas used for the canonicalization cost evaluation center on the education-domain Open Badges v3.0 profile; generalization to production schemas in other domains (multi-namespace mDL mdocs, EUDIW PID/EAA schemas, and so on) requires further experiments. The measurements were performed with a Node.js implementation on four machines spanning x86\_64 and arm64 (Table~\ref{tab:4}), so absolute values may differ in other language runtimes, browser/WebCrypto environments, or on mobile devices. The structure of the attack surface---the presence or absence of canonicalization and external references---derives from the specifications and is therefore implementation-independent, but the degree to which it materializes depends on the implementation and on library configuration.

\textit{Validity of the security evaluation.} The security tests against W3C VCDM were conducted with default settings (no document loader restriction, no call limit, no context allowlist). One mitigation is quantified---Section~\ref{sec:r-limit} sweeps the call limit against both attack and legitimate input---but the combined effect of a production-recommended configuration was not measured. The tests against SD-JWT VC and mdoc target a limited set of attack vectors (alg:none, algorithm confusion, data element and COSE header tampering) and do not constitute a comprehensive security audit; moreover, as Section~\ref{sec:threat} notes, those verifiers were written for this study and follow the practices their specifications recommend, so their results characterize a conformant implementation rather than an arbitrary third-party one. The attack vectors evaluated are known threats documented in W3C specifications and the academic literature; zero-day vulnerabilities and novel attack techniques are out of scope.

\section{Conclusion}

This paper evaluated, at the implementation level and under an explicit threat model, the attack surface of the verification pipelines of SD-JWT VC, W3C VCDM, and mdoc, and quantified the mitigations that implementers should adopt.

For the JSON-LD/RDFC verification pipeline, we exercised all three structural attack surfaces arising from remote context resolution and RDFC-1.0 canonicalization under default settings that lack mitigations: the poison graph as a completed attack, context injection as an expansion redirected to a different IRI under a remotely resolved context without \texttt{@protected}, and SSRF as reachability at the document loader boundary. A poison graph of five blank nodes in 527 bytes raises canonicalization cost to 278$\times$ the baseline, and eight blank nodes in 1,196 bytes to 13.5 seconds (302$\times$ and 36.5 seconds in Environment~B, the bare-metal Celeron); the cost tracks the symmetry of the references among blank nodes rather than their number, so a bound on blank node count is the wrong control and a call limit on the recursion is the right one---at 16 or below it stops every attack family we built while leaving every legitimate credential tested unaffected, though it cannot constrain an input whose recursion is shallow and wide, such as the acyclic control. Context injection changes the IRI a term denotes when the context is resolved remotely and lacks \texttt{@protected}, and a recording document loader observes the verifier asking for the cloud instance metadata endpoint. Separately from all of this, adversarial input turns out not to be required for the cost concern at all: production schemas reach 23--34$\times$ a simple credential in Environment~A and 16--49$\times$ across the four environments with a growth rate that rises with attribute count. The effective mitigations are a static document loader, which removes the retrieval latency and both the injection and SSRF surfaces together; a context allowlist; and a call limit for RDFC-1.0 combined with a bound on document size.

By contrast, within the signing pipeline evaluated here SD-JWT VC and mdoc require neither canonicalization nor external references, and the specification-conformant verifiers we implemented for them rejected every attack vector evaluated---alg:none, algorithm confusion, data element tampering and COSE protected header tampering. As Section~\ref{sec:threat} states, those two results characterize what a conformant verifier does, not what an arbitrary third-party implementation does by default; the \texttt{alg:none} rejection in particular is a property of the algorithm allowlist the verifier is configured with. These are not format-specific defects but attack surfaces that depend on the verification pipeline and its operational configuration. We further argued, from the definition of the signing input in each cryptosuite, that protection of the meaning of a context depends on the securing mechanism: on RDFC-1.0 paths the expansion of the context is bound to the signature and a substitution is detected---which our test confirms---whereas on JCS and enveloping paths only the URL string is signed, so replacing the referenced document would pass verification. The latter case is argued rather than demonstrated here. Where JCS or enveloping protection is used, integrity protection of related resources should be treated as a requirement rather than an optional measure.

In format selection, the requirement for semantic interoperability and the resulting security and operational cost should be weighed explicitly. When W3C VCDM is adopted, we recommend implementing the mitigations shown in this paper---static document loader, context allowlist, and call limit---as defaults. A comparison of signature verification performance across the formats is reported in our performance evaluation \cite{ref19}. Our evaluation framework and test tools are published as open source \cite{ref26}; the attack inputs, the call-limit sweep and the loader conditions reported here are all suites of that kit, and we hope they support replication under different environments and conditions.

\section*{Data Availability}

The raw result files behind every number in this paper are deposited at \url{https://doi.org/10.5281/zenodo.23933321} under CC BY 4.0. The deposit holds eight runs---the four environments of Table~\ref{tab:4}, each measured once with the attack-surface suites reported here and once with the performance suites of the companion paper \cite{ref19}. Each run carries its own \texttt{environment.txt} recording the host, the isolation settings and the kit commit, a \texttt{summary.json} giving every condition's runs, n, mean, standard deviation, 95\% confidence interval and percentiles, and the five independent per-run outputs those statistics are computed from. The per-condition counts that the wall-clock budget reduced, noted beneath Table~\ref{tab:7}, are recorded there per run. The accompanying README maps runs to environments and gives the \texttt{jq} commands that read the figures reported here back out of the archive. The measurement kit that produced the files is separate and is cited as \cite{ref26}.

\section*{Appendix A: Security Test Implementation Code}

This appendix presents the code that produces the results of Sections~\ref{sec:r-tests} and~\ref{sec:r-poison}. Both listings are excerpts from the \texttt{poison} and \texttt{security} suites of vcbench \cite{ref26}; comments in the listings are those of the source.

\Needspace*{5\baselineskip}
\textbf{Listing 2. Poison Graph Families (\texttt{poison} suite)}\nopagebreak

\begin{lstlisting}
// Every node carries an explicit blank node
// identifier, so the references below really do
// form cycles. (A generator that omits @id yields
// a forest of disjoint trees, which is not a
// worst case for RDFC-1.0.)
const node = (i, targets) => ({
  '@id': `_:b${i}`,
  'http://example.org/link':
    targets.map((j) => ({ '@id': `_:b${j}` })),
})
const FAMILIES = {
  // K_n: every blank node references every other
  complete: (n) => ({ '@graph':
    Array.from({ length: n }, (_, i) => node(i,
      Array.from({ length: n }, (_, j) => j)
        .filter((j) => j !== i))) }),
  // bidirectional ring C_n
  ring: (n) => ({ '@graph':
    Array.from({ length: n }, (_, i) =>
      node(i, [(i + 1) % n, (i + n - 1) % n])) }),
  // 3-regular circulant graph
  cubic: (n) => ({ '@graph':
    Array.from({ length: n }, (_, i) => node(i,
      [(i + 1) % n, (i + n - 1) % n,
       (i + Math.floor(n / 2)) % n])) }),
  // acyclic forest from a generator omitting @id
  // (kept as a control)
  forest: (n) => ({ '@graph':
    Array.from({ length: n }, (_, i) => [
      { '@type': 'http://example.org/Node',
        'http://example.org/link':
          { '@id': `_:b${(i + 1) % n}` } },
      { '@type': 'http://example.org/Node',
        'http://example.org/link':
          { '@id': `_:b${i}` } },
    ]).flat() }),
}

// Input shapes are reported so that "n" is
// unambiguous: the number of blank nodes after
// expansion is not always the generator's
// parameter.
async function shapeOf(doc) {
  const nq = await jsonld.normalize(doc, BASE)
  const lines = nq.trim().split('\n').filter(Boolean)
  const bn = new Set()
  for (const l of lines)
    for (const m of l.matchAll(/_:c14n\d+/g))
      bn.add(m[0])
  const subj = new Set(lines
    .map((l) => l.split(' ')[0])
    .filter((x) => x.startsWith('_:')))
  const objs = new Set(lines
    .map((l) => l.split(' ')[2])
    .filter((x) => x.startsWith('_:')))
  return { quads: lines.length, blankNodes: bn.size,
    cyclic: [...subj].filter((x) => objs.has(x)).length > 0,
    bytes: Buffer.byteLength(JSON.stringify(doc)) }
}

// call-limit sweep (rdf-canonize maxDeepIterations)
for (const [label, { doc, opts }] of
     Object.entries(CASES)) {
  for (const k of [1, 4, 16, 64]) {
    await measure(`poison/limit-k${k}/${label}`,
      doc, { ...opts, maxDeepIterations: k }, 100)
  }
}
\end{lstlisting}

\Needspace*{5\baselineskip}
\textbf{Listing 3. SSRF Reachability and mdoc Tampering (\texttt{security} suite)}\nopagebreak

One simplification in the mdoc half of this listing should be stated. ISO/IEC 18013-5 computes each value digest over \texttt{IssuerSignedItemBytes}, the \texttt{IssuerSignedItem} map wrapped in a CBOR tag-24 byte string; the harness digests the untagged encoding instead. Because the same encoding is used when the MSO is built and when it is checked, the element-tampering and header-tampering verdicts are unaffected, but the digests are not byte-compatible with a conformant mdoc.

\begin{lstlisting}
// A recording document loader observes the request
// at the loader boundary without any packet leaving
// the process, so reachability is established by
// observation rather than by reading the source.
const requested = []
const recordingLoader = async (url) => {
  requested.push(url)
  throw new Error(`blocked: ${url}`)
}
const evil = { '@context':
  ['http://169.254.169.254/latest/meta-data/'],
  '@id': 'https://example.com/c/1' }
try {
  await jsonld.normalize(evil,
    { ...OPTS, documentLoader: recordingLoader })
} catch { /* expected */ }
record('vcdm/ssrf-reachability',
  requested.length ? 'OBSERVED' : 'NOT-OBSERVED',
  `loader was asked for: ${requested.join(', ')}`)

// mdoc: each element is digested individually and
// the digests are signed inside the MSO, so a
// modified element must fail digest comparison and
// a modified protected header must fail the
// COSE_Sign1 signature check.
const verify = (ph, elems, msoBytes, signature) => {
  if (!crypto.verify('sha256',
      sigStruct(ph, msoBytes),
      { key: publicKey, dsaEncoding: 'ieee-p1363' },
      signature)) {
    throw new Error(
      'COSE_Sign1 signature verification failed')
  }
  const digests = cborDecode(msoBytes)
    .valueDigests['org.iso.18013.5.1']
  for (const enc of elems) {
    const it = cborDecode(enc)
    if (!Buffer.from(digests[it.digestID])
        .equals(sha256(Buffer.from(enc))))
      throw new Error('digest mismatch')
  }
  return true
}
const tampered = [...encoded]
tampered[0] = cborEncode(
  { ...cborDecode(encoded[0]), elementValue: 'Attacker' })
try {
  verify(protectedHdr, tampered, mso, sig)
  record('mdoc/element-tamper', 'ACCEPTED',
    'modified element value was not detected')
} catch (e) {
  record('mdoc/element-tamper', 'REJECTED', e.message)
}
\end{lstlisting}

\balance

\end{document}